\documentclass[pdflatex,sn-nature]{sn-jnl}

\usepackage{graphicx}%
\usepackage{multirow}%
\usepackage{amsmath,amssymb,amsfonts}%
\usepackage{amsthm}%
\usepackage{mathrsfs}%
\usepackage[title]{appendix}%
\usepackage{xcolor}%
\usepackage{textcomp}%
\usepackage{manyfoot}%
\usepackage{booktabs}%
\usepackage{algorithm}%
\usepackage{algorithmicx}%
\usepackage{algpseudocode}%
\usepackage{listings}%
\usepackage{longtable}
\usepackage{booktabs}
\usepackage{array}
\usepackage{ragged2e}

\theoremstyle{thmstyleone}%
\theoremstyle{thmstyletwo}%

\theoremstyle{thmstylethree}%

\begin{document}

\title[Article Title]{Persona-Based Simulation of Human Opinion at Population Scale}


\author*[1]{\fnm{Mao} \sur{Li}}\email{maolee@umich.edu}

\author[1]{\fnm{Frederick G.} \sur{Conrad}}\email{fconrad@umich.edu}

\affil[1]{\orgdiv{Institute for Social Research}, \orgname{University of Michigan}, \orgaddress{\street{426 Thompson St.}, \city{Ann Arbor}, \postcode{48104}, \state{MI}, \country{US}}}


\abstract{What does it mean to model a person, not merely to predict isolated responses, preferences, or behaviors, but to simulate how an individual interprets events, forms opinions, makes judgments, and acts consistently across contexts? This question matters because social science requires not only observing and predicting human outcomes, but also simulating interventions and their consequences. Although large language models (LLMs) can generate human-like answers, most existing approaches remain predictive, relying on demographic correlations rather than representations of individuals themselves.

We introduce SPIRIT (Semi-structured Persona Inference and Reasoning for Individualized Trajectories), a framework designed explicitly for simulation rather than prediction. SPIRIT infers psychologically grounded, semi-structured personas from public social  media posts, integrating structured attributes (e.g., personality traits and world beliefs) with unstructured narrative text reflecting values and lived experience. These personas prompt LLM-based agents to act as specific individuals when answering survey questions or responding to events.

Using the Ipsos KnowledgePanel, a nationally representative probability sample of U.S. adults, we show that SPIRIT-conditioned simulations recover self-reported responses more faithfully than demographic persona and reproduce human-like heterogeneity in response patterns. We further demonstrate that persona banks can function as virtual respondent panels for studying both stable attitudes and time-sensitive public opinion.}




\maketitle
\section{Introduction}

Large language models (LLMs) are increasingly used to simulate opinion formation, opinion change, and decision making, opening new possibilities for computational social 
science and user-centric NLP applications 
\citep{argyle_out_2023,horton_large_2023,park_generative_2024}. 
A common approach gives the model a short demographic profile (for example age, gender, education), often called a persona, and asks it to answer as if it were a person with that profile.
\citep{argyle_out_2023,horton_large_2023}. 
This approach is conceptually appealing because demographics are widely available 
and central to population inference.  However, it assumes that demographics alone largely account for variation in opinions and behaviors across people. Recent studies show that results can change sharply when the question wording, examples, or formatting in the prompt changes.
\citep{wang_large_2025,murthy_one_2025,dillion_can_2023}.

In response, researchers have attempted to address these limitations in two main ways. One tries to increase diversity by simulating many more synthetic respondents, sometimes with a larger set of profile fields \citep{yang_oasis_2025,ge_scaling_2025,park_generative_2024}. 
These methods can produce a wider range of answers and may reduce obvious skews, such as always producing the same view for a given demographic group. But they still do not show that the mix of simulated people matches any real population, or that simulated group differences align with observed group differences.

Another approach focuses on enriching personas using detailed survey responses or text produced by the simulated individual, such as life narratives or interviews.
\citep{zhang_personalizing_2018,park_generative_2024,kang_deep_2025,moon_virtual_2024, toubia_twin-2k-500_2025}. 
These methods yield more vivid and coherent simulated responses, improving realism and consistency. Yet, they remain centered on individual-level fidelity 
and leave open the question of how collections of such personas relate to 
population-level distributions.

Despite their differences, both approaches leave unresolved a core challenge for using LLMs in social scientific inference: how to estimate population-level distributions, especially when personas are derived from either prompting an LLM to play a particular role (e.g., ``You are a 50-year-old Republican female from Kansas'') or sampling multiple users from non-probability sources, such as social media.
Scaling the number of synthetic respondents or enriching persona descriptions can make simulations look more realistic, but these approaches do not, by themselves, make the results suitable for population inference. The central problem is that the data obtained from the persona(s) are constrained by who they represent, which will not necessarily be the target population (often the general public). This mirrors a basic distinction in survey research: probability samples, with known inclusion probabilities, support design-based population estimation, whereas non-probability samples require additional assumptions and methods, such as calibration, to adjust for selection bias and unequal inclusion \citep{lohr_sampling_2021}. 

At a more fundamental level, demographics capture only a small portion of what shapes opinions and decisions \citep{liSimulatingSocietyRequires2025}. Many influential factors are not explained by age, gender, or education, such as personality traits \citep{mairesse_using_2007}, basic beliefs about the world \citep{clifton_primal_2019}, political identity \citep{jost_political_2012}, narrative identity \citep{mcadamsNarrativeIdentity2011}, and the information environments people inhabit \citep{schwartz_personality_2013}. As a result, demographic-based personas are often not sufficiently nuanced to provide realistic, person-specific responses. When the prompt underspecifies the individual person, the model tends to fill in the gaps using broad patterns learned during training \citep{ferrara_should_2023,mohsin_fundamental_2025}, which can dominate the simulated responses. The gap between what demographics determine about an individual's opinion and what the model must infer based on the particular set of demographics, limits the accuracy of persona-based simulations of public opinion and behavior at a population level. These limitations suggest that realistic simulation requires richer, more psychologically grounded person-representations than demographics alone can provide.

Studies of everyday self-expression on social media show that people’s language and behavior patterns are associated with self-reported personality measures, indicating that online traces reflect enduring differences between individuals \citep{bailey_authentic_2020}. Classic work by Kosinski and colleagues further showed that simple digital records, such as Facebook “Likes,” can be used to predict a range of personal attributes \citep{kosinski_private_2013}. Building on these findings, studies report that LLMs can infer traits directly from users’ social media text in a prompt-based setting, with performance comparable to models trained on labeled data in some tasks \citep{park_diminished_2023}. Evidence also suggests that this signal is not limited to social media: Wright and colleagues show that generative models can estimate Big Five personality traits (a common five-dimension personality framework) from brief open-ended narratives, producing respectable agreement with self-reports \citep{wright_assessing_2026}.

Prior work often stops at trait prediction and does not provide a clear way to organize inferred attributes into a reusable persona for simulation and evaluation. To move the enterprise forward, we introduce \textbf{SPIRIT} (\textbf{S}emi-structured \textbf{P}ersona \textbf{I}nference and \textbf{R}easoning for \textbf{I}ndividualized \textbf{T}rajectories), a framework that constructs structured, multi-dimensional personas from social media text and applies them in LLM-based behavioral simulation.

We first recruited panel members from a probability-based online panel (the Ipsos KnowledgePanel) who were also social media users, and collected their user handles on Reddit, Twitter, or both. Participants consented to linking their public posts to their Ipsos survey responses, which allowed the model to infer a SPIRIT persona for each individual. Because KnowledgePanel is designed to represent the U.S. population under a probability-based recruitment design, this starting point supports population-level inference.

We refer to the resulting collection of SPIRIT personas as a \textbf{Persona Bank}. Conceptually, it functions as a virtual twin panel that can be surveyed using standardized instruments. At the same time, our linkage procedure imposes additional eligibility requirements (having an account, posting, and consenting to link), which can introduce selection bias. We therefore construct weights and apply them to each persona-bank respondent to align weighted estimates with U.S. population benchmarks. By combining established ideas from survey methodology \citep{kish_survey_1995,little_statistical_2020,sarndal_model_2003} with LLM-based persona inference and simulation, SPIRIT provides a practical foundation for population-level simulation.

In summary, SPIRIT provides a persona framework in which personas are (1) inferred from authentic social media posts and (2) reweighted to support population-level generalization. Unlike prior approaches that focus on either increasing the number of synthetic respondents or enriching persona detail without addressing the sampling problem, SPIRIT operates at two levels: at the individual level, it prompts an LLM to adopt a person-specific persona to answer survey questions; at the population level, it combines those person-level responses with calibration weights so that aggregates can be interpreted as U.S. population estimates.

Across multiple survey questions on prominent political issues, SPIRIT persona outperforms demographic persona, producing simulations that are more stable across questions, easier to interpret, and more consistent at both the individual and population levels.

\textbf{Our contributions are threefold:}
\begin{enumerate}
    \item We introduce \textbf{SPIRIT}, a semi-structured persona framework that
    infers multi-dimensional user representations from social media text that are interpretable
    and uses them to prompt an LLM to respond to survey questions as the individual represented by the persona would respond.

    \item We provide a systematic evaluation of our framework, showing
    that demographic personas yield unrealistic response distributions and lower-confidence
    estimates, while \textbf{SPIRIT} personas better recover coherent, person-specific response
    patterns---highlighting the limits of demographic attributes alone for simulating how opinions are formed and change.

    \item We develop and empirically evaluate the \textbf{Persona Bank}. In particular, we calibrate persona-based simulations to population benchmarks and validate the resulting virtual panel against survey and polling results concerning salient political issues measured by high-quality opinion polls.
\end{enumerate}

\section{Results}\label{sec2}
The Ipsos KnowledgePanel collects a substantially broad set of respondent information; for the purposes of this study, we obtained a subset of 81 survey questions from the Ipsos records spanning a wide range of topics, from demographic characteristics to attitudes toward public health and vaccines. To construct a demographic persona for comparison, we use seven demographic variables—age, race, gender, political ideology, income, education, and urbanicity—as persona inputs. The remaining 52 non-demographic questions (spanning from self-rated health to political attitudes) are then used as held-out evaluation outcomes to assess how well different persona constructions recover respondents’ self-reported views and dispositions (full question wording is provided in Appendix~\ref{secA2}).

We evaluate the proposed framework along two complementary dimensions: (i) inference accuracy and (ii) the distribution of inferred user-level responses. We compare the \emph{demographic personas} against the \emph{SPIRIT personas}, which relies exclusively on non-demographic attributes inferred from user-generated text, across LLMs of increasing size.

\subsection{Framework Evaluation}
\begin{figure}[ht]
    \centering
    \includegraphics[width=\linewidth]{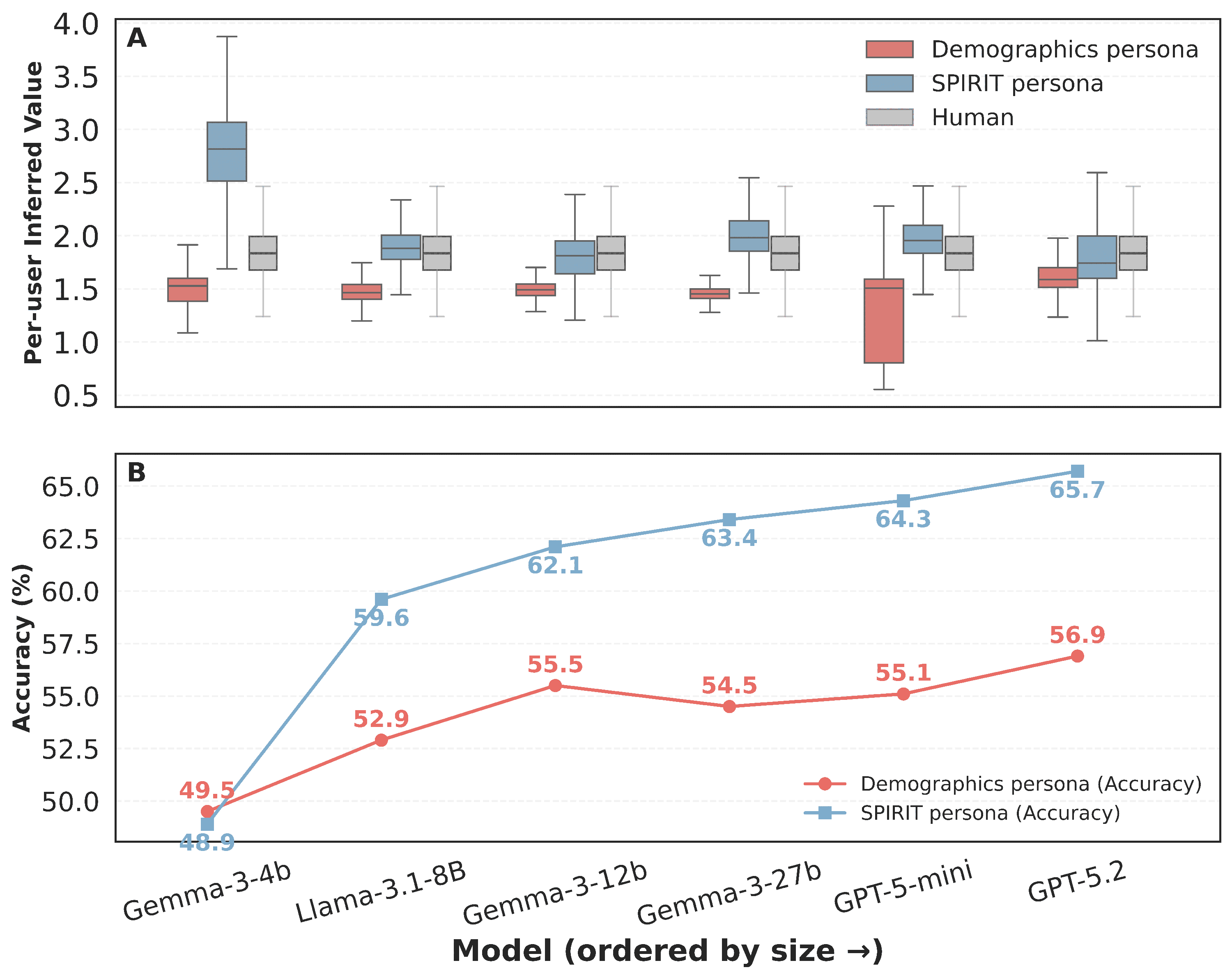}
\caption{
Framework evaluation across models and conditioning strategies.
\textbf{A}, Distribution of \emph{per-user position-weighted mean} inferred values for the same eligible Ipsos KnowledgePanel participants linked to public social media accounts, comparing their self-reported responses with simulated responses generated under the demographic persona and the SPIRIT persona (with non-demographic attributes inferred from text).
For each participant, responses are aggregated across survey items using a
position-weighted mean, such that identical composite values arise only from
identical response patterns.
Human responses are shown as an empirical reference for the level of
population-level heterogeneity expected when aggregating across many items.
Demographic personas yield highly concentrated distributions,
whereas SPIRIT personas preserve substantially greater
individual-level variation, closely resembling human heterogeneity.
\textbf{B}, User-level inference accuracy across models ordered by size.
SPIRIT personas consistently outperform demographic personas with performance gains saturating for larger models.
}
    \label{fig:framework-eval}
\end{figure}

\subsubsection{Accuracy of user-level inference}

Inference accuracy is evaluated at the user level.
Each user is associated with a set of questions; for each question, the model produces
an inferred response that is compared against the user’s self-reported value.
We compute the mean accuracy per user and then average across users within each
persona condition (i.e., SPIRIT persona vs. demographic persona).

As shown in Figure~\ref{fig:framework-eval}B,
under the SPIRIT personas, inference accuracy increases monotonically with model size.
Smaller models (e.g., Gemma-3-4B and LLaMA-3.1-8B) perform substantially worse than larger models,
while gains begin to plateau from Gemma-3-27B to GPT-5-mini and GPT-5.2.
This pattern is expected: the core requirements of the framework---reading user posts,
constructing non-demographic persona representations, and reasoning over them---rely
primarily on stable comprehension and moderate-level reasoning rather than highly complex capabilities (e.g., Coding or solving math problems).
Once these competencies are met, additional model capacity yields diminishing returns.

Comparing the demographic personas with the SPIRIT personas,
we observe a consistent accuracy advantage for the SPIRIT personas across all models.
While demographic personas' accuracy improves modestly with model size,
these gains presumably reflect improved guessing by larger models based on correlations between demographics and responses to survey items (e.g., income and credit score), rather than grounded inference from individual-specific signals.
Overall, the observed 8--9\% absolute improvement indicates the SPIRIT personas substantially outperform the demographic personas.

\subsubsection{Distributional properties of inferred responses}

Accuracy alone does not capture how well LLMs simulate the full pattern of user responses, especially their variability: accuracy can remain high even when inferred answers are overly concentrated.
We therefore examine the distribution of inferred user-level mean responses under each
persona condition.
For each user, we aggregate inferred responses across all 52 survey questions into a single composite score that summarizes their overall response pattern. We compute a \emph{position-weighted per-user mean}, where each item response is weighted by its position in the survey sequence (details are provided in Methods~\ref{sec:methods_position_weighted}).

A key property of this construction is that two users will obtain the same composite score only if their response sequences are identical across all questions. Because perfectly identical sequences are unlikely in realistic survey settings, and would effectively eliminate individual heterogeneity, we expect substantive user-level differences to manifest as a dispersed distribution of composite scores rather than one concentrated around a single value. This composite score preserves the ordering of responses while reducing each user’s response profile to a single summary measure that still captures meaningful heterogeneity across users.

The same aggregation rule is applied to inferred responses and human self-reports.
The resulting distributions are visualized using box plots
(Figure~\ref{fig:framework-eval}A).

To provide a meaningful reference point, we include the distribution of
human self-reported responses.
This human distribution is not treated as a target to be matched exactly,
but rather as an empirical benchmark for population-level heterogeneity.

Across all models, SPIRIT personas produce broader and more differentiated
distributions than demographic personas, more closely resembling the spread
observed in human responses.
In contrast, the demographic personas yield markedly narrower distributions,
indicating homogenized predictions driven by shared demographic profiles.

This behavior is expected.
For certain questions, such as credit score or media consumption, the demographic personas provide strong predictive signals. For example, high household income is strongly associated with excellent credit scores, and age correlates with lower social media use and greater reliance on television or newspapers. When such correlations dominate, demographic persona inference tends to collapse individuals with similar demographic profiles into near-identical predictions, thereby reducing population-level diversity.

Consistent with this pattern, the median inferred value under demographic personas is concentrated around approximately 1.5. This reflects a tendency for demographic persona inference to default to negative responses (e.g., No) for many behavioral and health-related questions (e.g., drinking alcohol or reporting anxiety-related conditions), which substantially compresses the range of possible inferred responses. In addition, the demographic personas yield lower confidence estimates than the SPIRIT personas, a result that is expected given the limited informational content available when inference relies solely on demographic attributes. Further analyses unpacking these distributional and confidence differences are provided in Appendix~\ref{secA3}.

SPIRIT personas mitigate this effect by incorporating linguistic, behavioral, and psychological signals extracted from user-generated text.
This is particularly salient for attributes weakly determined by demographics, including psychological traits and mental health-related indicators.
As a result, SPIRIT personas preserve individual-level variation that is otherwise lost under demographic personas.

It is important to note that inference accuracy is computed using a strict exact-match criterion between inferred values and self-reported responses. This metric does not account for measurement error inherent in survey responses. Prior work has shown that human responses—particularly for Likert-type items—are subject to instability over time \citep{park_generative_2024, alwin_reliability_1991}; even when the same individual is re-interviewed after a short interval, responses may differ due to interpretation, recall, or response scale ambiguity. For example, distinctions between categories such as sometimes and often may reflect measurement noise rather than substantive disagreement.

To account for this, we additionally compute an \emph{off-by-one rate} for the best-performing configurations (GPT-5.2 and GPT-5-mini with SPIRIT personas). The off-by-one rate is defined as the proportion of inferred responses whose ordinal distance from the corresponding self-reported value is exactly one category, aggregated across survey items at the individual level. We find off-by-one rates of 0.18 for GPT-5.2 and 0.19 for GPT-5-mini, indicating that approximately 83\% of inferred responses are either exact matches or differ by at most one response category from the self-reported value.

We focus our main comparison on the 52 non-demographic items because the remaining 29 items are demographic variables, and some of these are used to construct the demographic persona. Including them in the primary evaluation would therefore make the comparison less fair. We nevertheless extend the analysis to all 81 items as a supplementary check. Because SPIRIT is inferred only from users’ historical Reddit and Twitter/X posts and does not use any self-reported survey attributes, including demographics, this broader analysis allows us to assess whether the inferred personas can also recover demographic characteristics that were never given as inputs.

A clear pattern emerges across domains. SPIRIT performs particularly well on health- and vaccine-related items, which tend to reflect more stable beliefs and to show stronger consistency across related questions. In contrast, performance is weaker on short-horizon behavioral items such as beer consumption (e.g., whether someone drank in the past week or month). These behaviors are inherently more variable and sensitive to recent circumstances, so even the same person may answer differently over time, making them harder to infer reliably from long-run posting histories.

Importantly, this is not the primary use case we emphasize. Our goal is to support inference and simulation of comparatively stable orientations and belief systems, where internal consistency is expected and substantively meaningful, rather than to predict transient, week-to-week behaviors. The results of this extended evaluation are reported in Appendix~\ref{secA5}.

These results show that the proposed framework improves not only inference accuracy
but also the realism and expressiveness of inferred user responses.
While demographic personas perform well for easily predictable attributes
probably because population-level correlations, SPIRIT personas capture individual differences and avoid overly homogeneous simulations. 

\subsection{Persona banks as virtual respondent panels}

While the first part of the results establishes the validity of the proposed framework by benchmarking simulated responses against self-reported values, such validation alone is not the ultimate objective of this work.
The self-reported data available for benchmarking may not themselves be of substantive interest to researchers. The broader goal is to enable SPIRIT personas to simulate public opinion across a wide, if not unconstrained, range of topics and questions.

We therefore turn to a second set of analyses that treat the group of SPIRIT personas as a \emph{persona bank}, i.e., a collection of virtual respondents that can be surveyed using standardized survey instruments. Conceptually, this persona bank serves as a virtual respondent panel, allowing us to examine public attitudes toward rapidly evolving and event-driven issues for which traditional survey data are often delayed or expensive.

\subsubsection{Surveying the persona bank}

We focus on four public issues that were highly salient during the study period:
(i) opinions about abortion \citep{pew_research_center_public_2025},
(ii) attitudes toward immigration \citep{krogstad_trump_2024}, 
(iii) public reactions to the Epstein files \citep{taylor_orth_bipartisan_nodate}, and
(iv) views on U.S. military actions in Venezuela \citep{anthony_salvanto_cbs_2025}.
For each topic, we reused survey questions from contemporaneous public opinion polls, preserving the original wording and response options.
These questions are posed to the persona bank, and the resulting simulated responses are aggregated to produce population-level estimates, with each respondent weighted accordingly (as described in Method~\ref{sec:weights}). Full question wordings and response options are provided in Appendix~\ref{secA4}.

We interpret the first two question clusters (i.e., opinions about abortion and attitudes toward immigration) as measuring relatively stable, crystallized opinions that tend to be anchored in enduring belief structures \citep{converseNatureBeliefSystems2006}.
By contrast, clusters (iii) and (iv) capture more time- and event-sensitive judgments, for which expressed opinions are often constructed from whatever considerations and facts are most salient at the moment of response \citep{zaller_simple_1992}. Accordingly, for clusters (iii) and (iv), we allow simulated LLM respondents to retrieve up-to-date external information via web search, as described in Method~\ref{sec:methods_external_survey}.


\subsubsection{Trend alignment with polling benchmarks}

\begin{figure}[t]
    \centering
    \includegraphics[width=\linewidth]{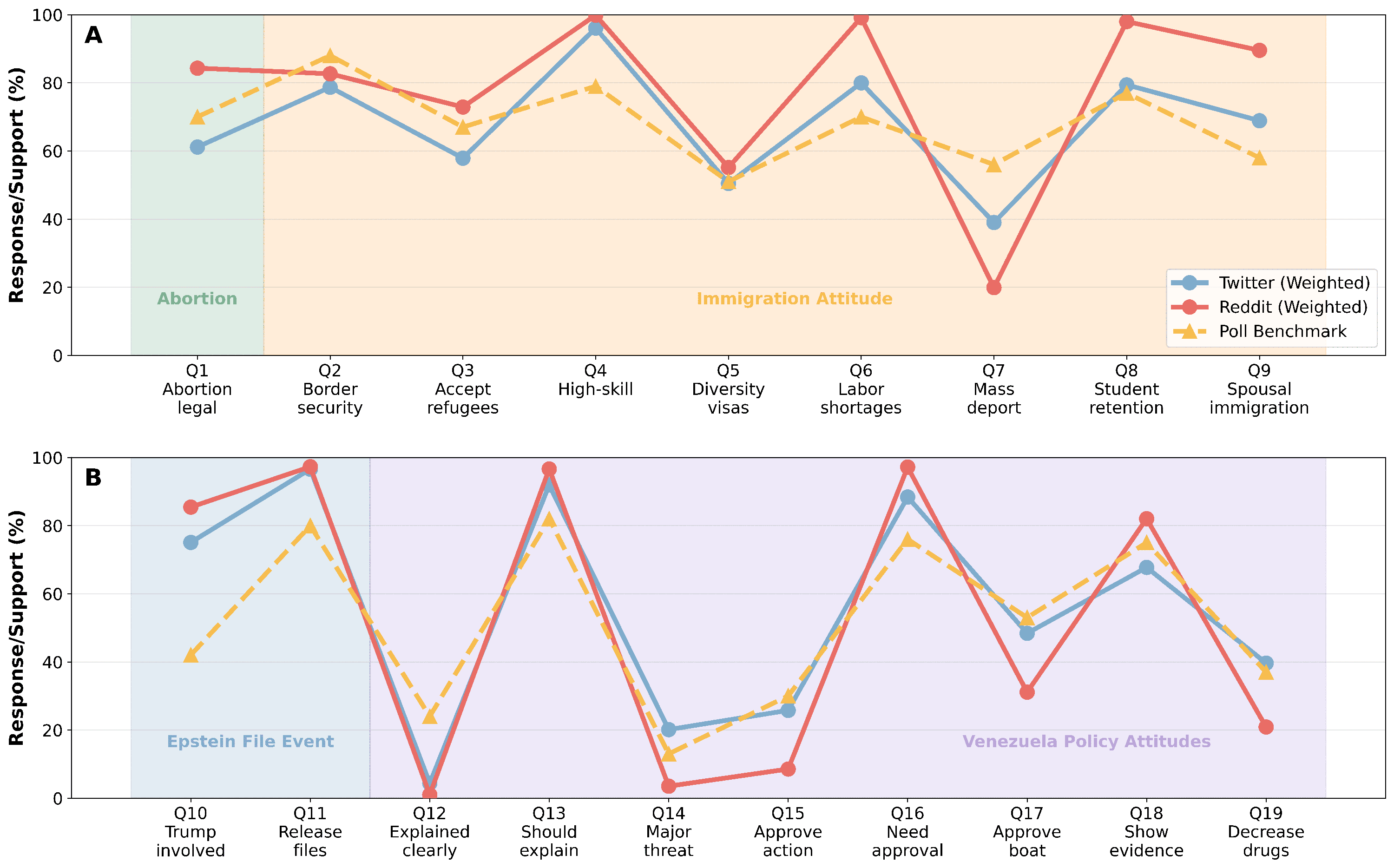}
    \caption{
    persona-bank responses compared with polling benchmarks, grouped by question type.
    \textbf{A}, Long-term attitudinal questions (abortion and immigration) drawn from general opinion surveys.
    \textbf{B}, Event-sensitive questions (Epstein files and Venezuela policy attitudes) fielded in late 2025 to early 2026,
    for which simulated respondents may require contemporaneous context.
    Shaded regions indicate issue-specific clusters and are used to avoid implying continuity across unrelated issues.
    Within each issue-specific cluster, persona-bank estimates reproduce coherent question-to-question patterns that align with polling benchmarks.
    After calibration, Twitter-based estimates track benchmarks more closely in absolute level than Reddit-based estimates, which exhibit systematic shifts in magnitude. Both Twitter- and Reddit-based estimates preserve the same directional structure.
    }
    \label{fig:poll_alignment}
\end{figure}

Figure~\ref{fig:poll_alignment} compares weighted estimates derived from Twitter- and
Reddit-based SPIRIT persona banks with polling benchmarks across multiple domains.
All simulated responses are generated by querying the same underlying set of
simulated respondents, i.e., the persona-bank, which makes comparisons across questions within an issue-specific cluster directly interpretable.
In contrast, the \textit{polling} benchmarks are drawn from different survey organizations
and respondent samples, and therefore should not be interpreted as forming a
single, internally comparable scale across clusters.

Accordingly, our evaluation emphasizes within-cluster pattern alignment rather than
cross-cluster comparisons.
We focus on whether persona-bank responses move up as polling benchmark estimates move up and down as polling benchmark estimates move down across related questions within each issue-specific cluster.

In Panel~A, which displays measures of long-term and crystallized attitudes on abortion and immigration,
the Twitter-based persona bank closely tracks polling benchmarks after calibration,
both in magnitude and in direction of question-to-question changes.
The Reddit-based persona bank exhibits larger shifts in absolute levels, particularly
for immigration-related items, but preserves consistent directional patterns that
mirror the polling benchmark.
As documented in Appendix~\ref{app:demographics}, these differences in magnitude are consistent with known compositional differences in the Reddit sample, which is more educated and more politically liberal than the population in general.

Panel~B examines event- and time-sensitive questions related to the Epstein files and U.S. policy toward Venezuela.
Despite the additional complexity introduced by time-sensitive information and heterogeneous polling sources, persona-bank responses again reproduce coherent
within-cluster structure.
Items eliciting stronger support in polling data are generally ranked higher by
persona-bank responses, while lower-support items remain comparatively low.
This agreement across polling references strengthens the interpretation that persona-bank responses capture coherent, item-to-item attitude patterns (that is, consistent changes in direction across related questions).

Importantly, these trend-alignment patterns are not an artifact of calibration.
When responses are aggregated without weighting, persona-bank estimates continue to
reproduce the same within-cluster directional structure, albeit with substantially
larger deviations in absolute levels. Weighting mainly reduces the magnitude of these differences, which we attributed to the slection bias, i.e., who is and who is not represented in the sample. Unweighted results are reported in Appendix~\ref{app:unweighted}, showing that weighting improves accuracy without changing the main conclusions.


At the same time, we observe two systematic deviations from typical human survey behavior.
For the items asking ``whether the Epstein files should be released'', ``Has the Trump administration clearly explained what the U.S. intends to do regarding Venezuela?'' or ``Does the Trump administration need to explain...'' nearly all simulated
respondents selected ``Yes,'' ``has not explained clearly'' and ``needs to explain'', whereas attitudes from real-world respondents are more heterogeneous.
Survey respondents may answer these questions based on more nuanced consideration, e.g., affect, partisan cues, or general
distrust in institutions, rather than treating them as relatively factual judgments, the way LLM seems to do.

Another deviation arises for items that implicitly require respondents to reason through a multi-step mechanism. Consider: “Do you think U.S. military action in Venezuela would decrease the amount of drugs coming into the U.S.?” In conventional survey settings, many respondents do not fully account for the link between policy interventions and downstream outcomes. Instead, they often rely on low-effort heuristics: a response of “Yes, would decrease” can function as an intuitive, affirmation of the policy, whereas “no effect” can express doubt about the policy with low effort and without having to explain why. In contrast, simulated agents are systematically more likely to “work the problem.” They lay out a causal chain (e.g., displacement of routes, adaptation by trafficking organizations, enforcement fragmentation, and downstream supply responses) and, after tracing these steps, frequently converge on “increase” as the most coherent implication of intervention. This pattern is consistent with established evidence that survey respondents often satisfice rather than optimize when questions are cognitively demanding \citep{krosnickResponseStrategiesCoping1991}, whereas the simulated agents are not programmed to reduce effort. These deviations are therefore not random errors, but reflect a systematic difference in how human and simulated agents respond to surveys.
Additional topic- and item-level analyses of these patterns are reported in
Appendix~\ref{app:venezuela_deviation}.

Taken together, these results indicate that persona banks can serve as a credible source of information about public opinion.
Even when absolute levels diverge, most notably for Reddit, the SPIRIT personas reproduce coherent patterns of opinion and consistent trends.
This supports our intuition that persona-based virtual panels (i.e., the persona bank) are useful for comparative analysis and rapid-response measurement,
especially in settings where traditional surveys are expensive or too slow to field.

\section{Discussion}
This study introduces \textsc{SPIRIT}, a semi-structured persona inference framework that builds rich representations of individuals based on their attributes, inferred from their social media traces, extending well beyond demographics.
Because \textsc{SPIRIT} is built from an address-based probability panel, the target population is well-defined (i.e., US population), and we are able to provide calibration weights to project aggregated results back to U.S. population benchmarks.
Across two complementary evaluations, we demonstrate both the validity of the proposed framework
and the broader potential of surveying a reusable persona bank for downstream computational social science analyses.

\subsection{From validation to simulation}

In the first part of our analysis, we show that \textsc{SPIRIT} personas can
faithfully infers a wide range of self-reported attributes under a strict
exact-match evaluation.
These results establish the internal validity of the framework:
inferred SPIRIT personas are not arbitrary abstractions, but capture
meaningful individual-level signals reflected in survey responses.

However, recovering self-reported values is not the ultimate objective of
persona-based modeling.
Self-reports are inherently limited in scope and timeliness,
and many substantive research questions concern attitudes toward
events that unfold faster than traditional surveys can capture.
The second part of our results, therefore, reframes inferred SPIRIT personas as a
\emph{persona bank}—a virtual respondent panel that can be queried using
standardized survey instruments to study emerging public opinion.
Our findings show that, after appropriate weighting, persona-bank responses
reproduce trends across questions and topics that closely align
with contemporaneous polling benchmarks.
This suggests that persona-based virtual panels can serve as a
useful source for rapid-response ( e.g., immediately after a political debate or in the aftermath of a natural disaster) and exploratory public opinion research.

\subsection{Why demographic attributes alone are insufficient}

A central insight of this work is that simulating human opinion requires substantially more information than demographic attributes alone. Although demographics are an important part of who people are, our results show that personas comprised only of demographic attributes tend to produce compressed response distributions (i.e., simulated answers concentrate on a narrow subset of options) and answers that lean heavily on broad demographic ``stereotypes'' rather than on person-specific evidence.

Importantly, in our experimental design, we intentionally exclude demographic attributes from the \textsc{SPIRIT} personas in order to isolate the contribution of non-demographic signals derived from users’ historical posts. This design choice should not be read as a recommendation to omit demographics in applied settings. On the contrary, \textsc{SPIRIT} is designed to accommodate as much relevant information as possible. The broader goal is to construct personas that are richly grounded in lived experience, so the model reasons not about what a typical person with certain demographics might think, but about how a particular individual who has a particular demographic profile would plausibly respond given their expressed values, beliefs, and behavioral patterns. In short, demographics are a core component of identity, but they are not the whole story.

\subsection{The role of data richness and user expression}

The quality and richness of input data play a critical role in the success of
persona-based inference.
In our study, users whose attitudes are most faithfully reconstructed are
typically those who leave dense and expressive digital traces,
sharing opinions, experiences, and reactions across a variety of contexts.
This observation highlights an important boundary condition: persona-based simulation is most effective when users provide meaningful signals in their language.

At the same time, this limitation also points to future opportunities.
Social media posts are only one source of qualitative information.
In principle, personas could be augmented with additional data sources,
such as open-ended and qualitative survey responses, diaries, or product reviews.
Extending persona construction beyond social media represents a promising
direction for building more comprehensive and structured representations
of individuals.

\subsection{Opinion questions, factual judgments, and model guardrails}

While \textsc{SPIRIT} performs well on many attitudinal questions, we also identify cases where it diverges from survey benchmarks. In particular, two types of items are more prone to mismatch. First, the model is more likely to diverge on items that go beyond subjective opinion and instead ask respondents to make evaluative judgments, especially when those judgments carry a strong moral or quasi-factual character. On these items, the simulated responses tend to be highly skewed in ways that do not match the observed survey distributions. Second, for questions that ask respondents to anticipate downstream consequences and therefore require causal reasoning, the simulated responses tend to be more consistently extreme and directional than those observed in human survey data.

These patterns suggest that large language models do not always act purely
as simulators of individual belief states.
Instead, they may default to responses shaped by \textit{built-in} safety constraints, normative assumptions, or implicit notions of ``correct'' behavior.
In these cases, the model behaves less like a representation of a specific
human respondent and more like a helpful assistant, optimized to produce
socially acceptable or morally justified answers.

This distinction highlights a fundamental tension in persona-based simulation.
For research purposes, the goal is not to predict what a population should think,
nor to output normatively desirable responses, but to approximate how individuals
actually think and react.
Even when explicitly instructed to ``be'' a particular person,
LLMs may override a persona in situations involving
strong moral priors or factual judgments.
Future work should examine how to represent the internal contradictions that characterize individuals, including socially undesirable traits, while also accounting for the built-in constraints that LLMs impose on harmful speech and behavior.

\subsection{Extending persona banks through agentic capabilities}

Finally, our results point toward a broader research agenda in which
persona banks are embedded within agentic systems.
To enable simulation of responses to newly unfolding events,
we augment persona-based reasoning with selective information retrieval,
allowing agents to access contemporaneous context when necessary.
This design parallels recent advances in agent frameworks that integrate
search, tool use, and external observation \citep{yao_react_2023}.

More broadly, persona-based agents need not remain respondents.
Future work could explore multi-agent interactions in which personas
deliberate, exchange information, or influence one another,
offering a controlled environment for studying opinion formation,
polarization, and social dynamics.
When combined with probability-based weighting and structured personas,
such systems may provide a valuable experimental testbed for
population-level reasoning under well-defined assumptions.

\section{Methods}
\label{sec:methods}
\subsection{Data}

\subsubsection{Panel recruitment and consented account linkage}
We partnered with the \textit{Ipsos KnowledgePanel}, a nationally recruited, probability-based survey panel (i.e., whose members represent the U.S. population). Panelists were screened for active usage of \textit{Twitter/X} and \textit{Reddit} and invited to consent to the retrieval of their posts. Consenting participants provided their handles on either or both platforms, which enabled the collection of their publicly available posts and linkage to their survey responses. From consenting panelists, we received handles for 1,410 \textit{Twitter/X} users and 893 \textit{Reddit} users, including 452 individuals who provided accounts on both platforms.

Importantly, although panelists provided handles, not all of the submitted handles were valid or retrievable at the time of data collection. For example, some handles contained typographical errors, were suspended/deleted, or pointed to accounts that were private or otherwise inaccessible via the platform APIs. We therefore retained only handles that could be associated with (i) an account and (ii) publicly accessible content that could be retrieved through the APIs.

\subsubsection{Post collection and linkage to survey responses}
For each valid handle, we collected all publicly available posts retrievable via the platform's APIs, subject to platform access constraints at the time of collection. These posts were linked to their attitudes and preferences maintained by Ipsos, enabling the construction and evaluation of the individualized personas (Appendix~\ref{secA2}).

This design thus supports persona-based simulation evaluation by providing: (i) respondent-level ground truth for opinions and attitudes (via Ipsos survey responses collected independently of this work), (ii) longitudinal, historical text traces for persona construction (via all public posts), and (iii) a probability-based sampling frame with survey weights, enabling simulation fidelity to be assessed not only at the individual level but also in terms of population-calibrated aggregates benchmarked to the U.S. adult population.

\subsection{SPIRIT framework}
\begin{figure}[ht]
    \centering
    \includegraphics[width=\linewidth]{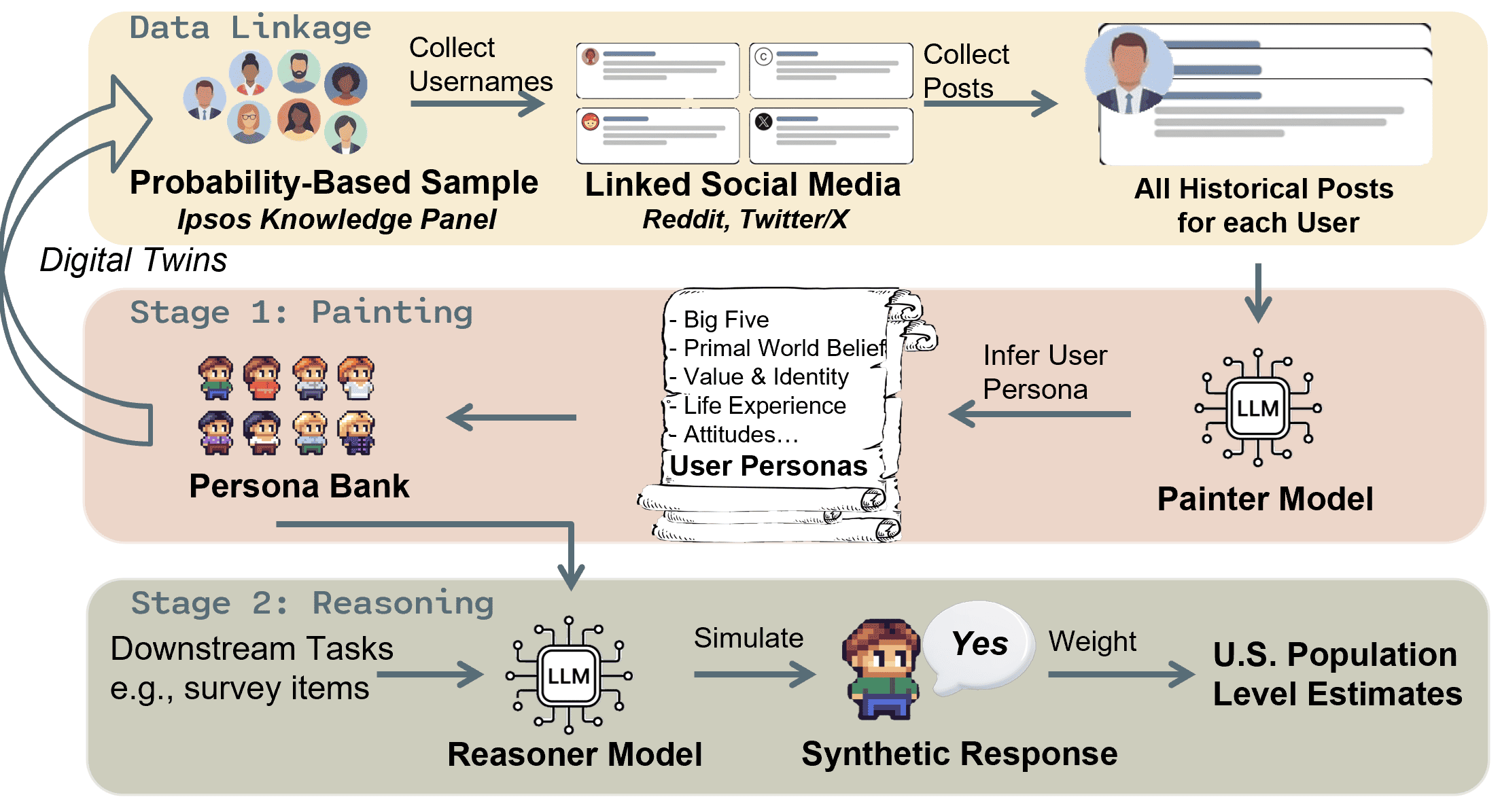}
\caption{
Overview of the SPIRIT framework. A probability-based sample from the Ipsos KnowledgePanel is linked to respondents’ social media accounts, and their historical posts are collected to infer structured user personas with a painter model. These inferred personas form a persona bank that serves as digital twins for Stage 2 reasoning, where a reasoner model simulates responses to downstream tasks such as survey items. The simulated responses are then weighted to produce U.S. population-level estimates.
}
    \label{fig:framework}
\end{figure}
\label{sec:methods_spirit_overview}
We propose \textbf{SPIRIT} (\textbf{S}emi-structured \textbf{P}ersona \textbf{I}nference and \textbf{R}easoning for \textbf{I}ndividualized \textbf{T}rajectories), a two-stage framework for grounding LLM-based simulation in interpretable, semi-structured, and faithful personas inferred from social media traces. As illustrated in Figure~\ref{fig:framework}, SPIRIT comprises (i) a \textbf{Painter} module that infers a multi-dimensional persona profile from a user’s social media posts, and (ii) a \textbf{Reasoner} module that prompts an LLM on the inferred persona to simulate structured responses for downstream tasks (e.g., survey-style items).

\subsubsection{Inputs and persona artifacts}
\label{sec:methods_inputs_artifacts}
For each user, the system consumes a single concatenated document consisting of all available historical posts from Reddit and Twitter/X, ordered by timestamp (each post is associated with its original posting time). The Painter outputs (detailed scheme can be seen in Appendix~\ref{app:persona_schema}):
(i) a \textbf{semi-structured persona profile} encoded as a JSON (JavaScript Object Notation) object that captures psychologically and socially meaningful attributes (e.g., traits, world beliefs, values/identities, attitudes), together with uncertainty annotations; and
(ii) a \textbf{narrative persona} that summarizes the profile in readable third-person form.

In external survey experiments, demographic attributes (e.g., age, gender, race, party identification) are loaded from user profiles and provided as supplementary content during downstream prompting.

\subsubsection{Painter: semi-structured persona inference from posts}
\label{sec:methods_painter}

\paragraph{Persona schema.}
The persona profile follows a fixed schema that is designed to be (i) multi-dimensional, covering traits (Big Five), worldview beliefs, values/identities, and domain attitudes, and (ii) auditable, with explicit uncertainty scores and brief justifications. The schema is implemented as a typed data model to enforce structural consistency across users.

\paragraph{Inference procedure.}
Given a user’s posts, the Painter infers persona attributes by synthesizing recurring linguistic signals (e.g., affect, self-references, moral language, expressed preferences, and stance patterns) into the persona schema. To reduce overinterpretation, Painter is instructed to treat all inferences as probabilistic and to express low confidence when evidence is weak. Prompt templates and an illustrative example are provided in Appendix~\ref{app:painter_prompt}.

\subsubsection{Reasoner: persona inference for downstream tasks}
\label{sec:methods_reasoner}

\paragraph{Task formulation.}
The Reasoner answers downstream questions by prompting an LLM with the inferred persona (and demographic attributes when available). For each question, the output includes: (i) a selected response option (value/label), (ii) a categorical confidence rating, and (iii) a brief textual rationale grounded in persona evidence. This formulation supports survey-style prediction enriched with qualitative evidence, as well as model uncertainty when persona signals are ambiguous.

\subsubsection{Calibration weights via raking to Census/ACS margins}
\label{sec:weights}

The persona bank requires respondents to (i) have a social media account, (ii) have ever posted, and (iii) consent to link. Because not everyone in the panel necessarily conforms to these requirements, the resulting sample might deviate from the U.S. adult population in basic composition.
To reduce this type of selection bias, we construct weights that calibrate the persona-bank marginals to external U.S. population benchmarks.

\paragraph{Benchmarks.}
We use marginal distributions from U.S. Census and ACS releases as targets:
gender (2020 Census), race/ethnicity (2020 Census), region (2020 Census), and age and education
(ACS 2022; education among adults 25+). We additionally include the panel variable
corresponding to the candidate for whom the respondent voted in the 2024 U.S. presidential election, using the election results (three-category vote preference) as a calibration margin.

\paragraph{Variables and preprocessing.}
Weights are computed for respondents included in the persona bank.
We rake on six margins: \texttt{PPGENDER}, \texttt{PPEDUC5}, \texttt{PPETHM}, \texttt{PPREG4},
\texttt{age\_group}, and \texttt{candidate2024}. Age is discretized into four groups
(18--29, 30--44, 45--64, 65+). 

\paragraph{Raking procedure.}
We apply iterative proportional fitting (raking) to match each sample margin to its target \citep{devilleGeneralizedRakingProcedures1993}.
Starting from equal base weights $w_i^{(0)}=1$, we iterate over margins and update weights
by multiplying an adjustment factor for the respondent's category:
\[
w_i \leftarrow w_i \times \frac{T_{v,c}}{\hat{P}_{v,c}},
\]
where $T_{v,c}$ is the benchmark proportion for category $c$ of variable $v$, and
$\hat{P}_{v,c}$ is the current weighted sample proportion for that category.
We cycle through all margins repeatedly until convergence (maximum absolute adjustment $tol<0.001$)
or a maximum of 50 iterations.

\paragraph{Normalization and use.}
After raking, we rescale weights to have mean 1 within the persona bank,
$w_i \leftarrow w_i/\bar{w}$, so weighted estimates preserve the sample's effective scale.
These weights are then used to compute weighted population-level
aggregates of simulated survey responses from the persona bank.
After normalization (mean $=1$), the raking weights show moderate dispersion:
$\text{SD}=1.32$, median $=0.63$ (IQR: 0.30--1.23), with a minimum of $0.012$
and a maximum of $14.42$ ($n=1517$).

\subsection{Position-weighted user-level response composite score}
\label{sec:methods_position_weighted}

To summarize each respondent’s overall response pattern across the survey, we convert
item-level outputs into a single user-level composite score. Let $i$ index users and
$q \in \{1,\dots,Q\}$ index survey items, where $Q=52$ in the main analysis. Each item
has an inferred numeric response $\hat{y}_{iq}$ (and, for human benchmarks, an observed
self-report $y_{iq}$).

We first assign each question a deterministic \emph{position weight} based on its
location in the survey sequence. In implementation, we take the set of unique
question IDs appearing in the analysis data, order them according to the survey
sequence used in the questionnaire, and assign weights $w_q \in \{1,\dots,Q\}$ so that
earlier items receive smaller weights and later items receive larger weights:
\begin{equation}
w_q = q.
\end{equation}
We then compute the position-weighted mean for each user:
\begin{equation}
\bar{\hat{y}}^{\,\mathrm{pos}}_i
=
\frac{\sum_{q=1}^{Q} w_q \, \hat{y}_{iq}}{\sum_{q=1}^{Q} w_q}.
\label{eq:pos_weighted_mean}
\end{equation}
We apply the same transformation to the human benchmark responses:
\begin{equation}
\bar{y}^{\,\mathrm{pos}}_i
=
\frac{\sum_{q=1}^{Q} w_q \, y_{iq}}{\sum_{q=1}^{Q} w_q}.
\end{equation}

This composite score is fully deterministic and model-agnostic. Because later items
receive larger weights, the summary retains information about the survey ordering while
yielding a compact scalar representation of each user’s 52-item response sequence. In the
empirical analysis, we compute $\bar{\hat{y}}^{\,\mathrm{pos}}_i$ for each model and experimental
condition and compare its distribution to the corresponding distribution of
$\bar{y}^{\,\mathrm{pos}}_i$ among human respondents.

\subsection{External survey stress tests with timely events}
\label{sec:methods_external_survey}

\paragraph{Motivation and topics.}
To evaluate whether persona-conditioned simulation can handle both stable attitudes and time-sensitive information needs, we constructed external survey item sets spanning four widely discussed topics at the time of writing: abortion, the Epstein files, U.S. actions related to Venezuela, and immigration policy. The Epstein-file and Venezuela-action conditions were treated as time-sensitive cases where events may plausibly post-date model training (i.e., will not be included in the training data).

\paragraph{Runtime stack.}
Generation is performed through a locally deployed, OpenAI-compatible \texttt{vLLM} endpoint, using \texttt{google/gemma-3-27b-it} as the base model. For time-sensitive conditions, the pipeline incorporates a web search component (i.e., Tavily, which is configured via a search backend) to obtain up-to-date context before answering.

\subsubsection{Time-sensitive protocol: persona-guided information acquisition}
\label{sec:methods_timely_protocol}
For time-sensitive topics, we use a lightweight information acquisition protocol. The model first elicits the persona’s pre-existing impressions, then generates persona-consistent search queries and summarizes retrieved information, and finally answers survey items conditioned on both persona and the synthesized context. This design separates \emph{prior beliefs} from \emph{newly acquired context}, allowing us to examine whether access to recent information changes inferred responses in a direction consistent with the persona.

\subsubsection{Stable-attitude protocol: direct persona-conditioned answers}
\label{sec:methods_stable_protocol}
For attitudes often treated as crystallized (e.g., on topics such as abortion and immigration policy), the model answers directly without external search, producing survey-style responses conditioned only on persona and demographics.

\paragraph{Appendix materials.}
Full prompt templates, schema specifications, and example model inputs/outputs for both Painter and Reasoner are provided in Appendix~\ref{app:persona_schema}.

\backmatter





\bmhead{Acknowledgements}

We thank the U.S. Census Bureau for supporting this research. We also thank Ipsos for providing access to survey data and for their support of the external benchmarking component.

\section*{Declarations}

\begin{itemize}
    \item \textbf{Funding.} The author(s) disclosed receipt of the following financial support for the research, authorship, and/or publication of this article: This work was supported by the U.S. Census Bureau under CB20ADR0160002.

    \item \textbf{Conflict of interest / Competing interests.} The authors declare no competing interests.

    \item \textbf{Ethics approval and consent to participate.} This study was conducted under IRB approval (HUM00259279), with informed consent from all participants. This study uses observational, user-generated content from public social media platforms (Reddit and Twitter/X) and does not involve any interaction or intervention with human participants. Because the content was not collected with consent for redistribution, we treat both the underlying posts and derived persona artifacts as sensitive research materials and adopt a data-minimization approach in dissemination.

    \item \textbf{Consent for publication.} We do not publish raw social media posts, usernames/handles, or direct identifiers. Any illustrative examples in the manuscript are paraphrased to reduce re-identification risk.

    \item \textbf{Data availability.} To balance transparency with privacy, we will release (i) aggregated results reported in the paper and (ii) data products for the external survey benchmarking component, including the survey question sets, response options, and item-level model outputs used for comparison. We do not publicly release the underlying raw social media posts or full persona artifacts (semi-structured persona JSON and narrative personas), as these may contain personally identifiable information or enable re-identification through linkage. Researchers who require access to persona artifacts for verification or extension may contact the corresponding author to discuss controlled-access arrangements.

    \item \textbf{Materials availability.} All non-sensitive study materials (e.g., external survey instruments, mapping tables, and evaluation scripts) will be made available with the public release of the external survey component. Materials that directly contain or reconstruct user-level social media traces will be withheld from public release.

    \item \textbf{Code availability.} Code for the SPIRIT pipeline (Painter and Reasoner modules), the external survey study, and the analysis scripts will be released upon acceptance of the manuscript. The release will include documentation sufficient to reproduce the reported external survey experiments using the publicly available materials, and guidance for running persona-based simulations under controlled data access.

\end{itemize}


\bigskip





\begin{appendices}

\section{Ipsos KnowledgePanel Participant Demographic Distribution}
\label{app:demographics}
This section details the demographic composition of the participants drawn from the Ipsos KnowledgePanel for the Twitter and Reddit arms of the study (see Table~\ref{tab:participant_demographics}). 

The sample distribution reflects several key characteristics of social media users that deviate from U.S. Census benchmarks. Notably, both samples are more \textbf{male-dominated} (approx. 61\%) and \textbf{highly educated} (over 50\% with a Bachelor's degree) than the general population. \textbf{Political ideology} is skewed toward the left, particularly on Reddit, where 58.8\% of participants identify as some level of Liberal. Additionally, the Reddit sample is significantly younger than the Twitter sample, with 76.1\% of respondents falling under the age of 45, compared to 56\% for Twitter.

\begin{table}[htbp]
\centering
\caption{Demographic Characteristics of Twitter and Reddit Participants}
\label{tab:participant_demographics}
\begin{tabular}{lcccc}
\toprule
 & \multicolumn{2}{c}{Twitter ($N=1,031$)} & \multicolumn{2}{c}{Reddit ($N=774$)} \\
\cmidrule(lr){2-3} \cmidrule(lr){4-5}
Characteristic & $n$ & \% & $n$ & \% \\
\midrule
\textbf{Gender} & & & & \\
\quad Male & 628 & 60.9 & 475 & 61.4 \\
\quad Female & 403 & 39.1 & 299 & 38.6 \\
\addlinespace
\textbf{Age Category} & & & & \\
\quad 18--29 & 192 & 18.6 & 197 & 25.5 \\
\quad 30--44 & 386 & 37.4 & 392 & 50.6 \\
\quad 45--60 & 290 & 28.1 & 135 & 17.4 \\
\quad 60+ & 163 & 15.8 & 50 & 6.5 \\
\addlinespace
\textbf{Race/Ethnicity} & & & & \\
\quad White, Non-Hispanic & 665 & 64.5 & 525 & 67.8 \\
\quad Hispanic & 146 & 14.2 & 108 & 14.0 \\
\quad Black, Non-Hispanic & 113 & 11.0 & 45 & 5.8 \\
\quad 2+ Races, Non-Hispanic & 54 & 5.2 & 46 & 5.9 \\
\quad Other, Non-Hispanic & 53 & 5.1 & 50 & 6.5 \\
\midrule
\textbf{Education} & & & & \\
\quad Bachelor's degree or higher & 545 & 52.9 & 427 & 55.2 \\
\quad Some college / Associate's & 285 & 27.6 & 205 & 26.5 \\
\quad HS Graduate / GED & 150 & 14.5 & 74 & 9.6 \\
\quad No HS diploma & 28 & 2.7 & 21 & 2.7 \\
\addlinespace
\textbf{Urbanicity} & & & & \\
\quad Urban & 430 & 41.7 & 346 & 44.7 \\
\quad Suburban & 443 & 43.0 & 330 & 42.6 \\
\quad Rural & 157 & 15.2 & 97 & 12.5 \\
\midrule
\textbf{Political Ideology} & & & & \\
\quad Liberal & 437 & 42.4 & 455 & 58.8 \\
\quad Moderate & 275 & 26.7 & 189 & 24.4 \\
\quad Conservative & 306 & 29.7 & 124 & 16.0 \\
\quad Refused & 13 & 1.3 & 6 & 0.8 \\
\bottomrule
\end{tabular}
\end{table}

\section{Experimental Environment and Computational Cost}\label{secA1}

\subsection{Computing Environment}

All framework-level experiments were conducted on a single-node, multi-GPU system equipped with \textbf{8 NVIDIA H100 GPUs (80GB memory each)}. Model inference was deployed using \textbf{vLLM} with \textbf{bfloat16 (BF16) precision}, enabling efficient batched inference while maintaining numerical stability for long-context inputs.

The system was configured to support long input sequences required for persona inference from social media histories. No gradient computation or model fine-tuning was performed; all experiments were conducted in inference-only mode.

\subsection{Token-Level Accounting}

To provide a transparent and model-agnostic estimate of computational usage, we report aggregated \emph{input} and \emph{output} token counts rather than wall-clock time or GPU-hours. This choice reflects the fact that inference cost is primarily driven by token volume and sequence length, while runtime depends on deployment-specific factors such as batching strategy and hardware utilization.

Table~\ref{tab:token_usage} summarizes total token usage across platforms for framework execution.

\begin{table}[h]
\centering
\caption{Aggregated token usage by platform}
\label{tab:token_usage}
\begin{tabular}{lrr}
\hline
Platform & Input tokens & Output tokens \\
\hline
Reddit   & 41{,}924{,}612 & 12{,}650{,}506 \\
Twitter  & 77{,}077{,}952 & 9{,}715{,}741  \\
\hline
\end{tabular}
\end{table}

Input tokens include persona inference prompts as well as persona-conditioned reasoning prompts. Output tokens correspond to structured persona representations and downstream generated responses.

\subsection{Cost Estimation and Variability}

We intentionally do not report GPU-hours or exact cost. Token-level accounting provides an order-of-magnitude indication of computational scale that remains comparable across hardware and deployment environments.

Actual runtime and cost may vary substantially depending on batching efficiency, concurrency, and hardware utilization. Moreover, the framework incorporates structured output validation using a schema-based mechanism implemented with \emph{Pydantic}, which introduces controlled variability in generation cost.

Specifically, model outputs are validated against a predefined JSON schema, with a retry mechanism capped at ten attempts per prompt. In practice, the number of retries required depends strongly on model capability: higher-capacity models exhibit substantially higher compliance rates with the schema and therefore require fewer retries. As a result, the reported token counts should be interpreted as coarse-grained estimates rather than exact execution costs.

These sources of variability reflect engineering trade-offs rather than conceptual limitations of the proposed framework.

\subsection{Practical Implications}

Despite these sources of variability, the overall computational footprint of the framework remains well within the reach of typical academic research environments. Because persona inference is performed once per user and then reused for multiple downstream tasks, the cost of persona construction can be amortized across analyses and scenarios. Importantly, the framework does not rely on model fine-tuning or large-scale retraining, making population-level experiments feasible without prohibitive computational overhead.

A key practical constraint is the reliability of persona construction. Although we implement a retry mechanism (up to ten attempts) to enforce schema-compliant outputs, smaller models (Gemma-3-4B and LLaMA-3.1-8B) still fail to produce valid persona representations for a nontrivial fraction of users. These failures propagate to downstream reasoning and reduce overall performance, suggesting a minimum model capacity requirement for robust deployment. Accordingly, we report results from these smaller models only as a reference point; they should not be interpreted as definitive evidence about the framework’s performance.

\section{Data Dictionary and Variable Definitions}\label{secA2}
This section details the complete set of survey questions collected from Ipsos KnowledgePanel and used in the evaluation (see Table~\ref{tab:survey_questions}). The table below maps the internal variable identifiers (e.g., pph10221) to the verbatim question text presented to participants and the available response options.
\begin{longtable}{>{\RaggedRight}p{3cm} >{\RaggedRight}p{6cm} >{\RaggedRight}p{6cm}}
\caption{Survey Questions and Response Options Dictionary} \label{tab:survey_questions} \\
\toprule
\textbf{Variable} & \textbf{Question Text} & \textbf{Response Options} \\
\midrule
\endfirsthead

\multicolumn{3}{c}%
{{\bfseries \tablename\ \thetable{} -- continued from previous page}} \\
\toprule
\textbf{Variable} & \textbf{Question Text} & \textbf{Response Options} \\
\midrule
\endhead

\midrule
\multicolumn{3}{r}{{Continued on next page}} \\
\bottomrule
\endfoot

\bottomrule
\endlastfoot

QFLAG & QFLAG & 1: Qualified \newline 2: Terminated \newline 3: Partial \newline 4: Non-responder \\
\addlinespace

pph10221 & Q37: Do you NOW smoke cigarettes? & 1: Every day \newline 2: Some days \newline 3: Not at all \\
\addlinespace

ppsi1916 & Q210: Do you currently use marijuana…? & 1: Every day \newline 2: Some days \newline 3: Not at all \\
\addlinespace

pph21906 & Q110: How often do you think vaccines have dangerous side effects? & 1: Never \newline 2: Rarely \newline 3: Sometimes \newline 4: Often \newline 5: Very often \\
\addlinespace

pph21908 & Q130: Overall do you think & 1: The benefits... outweigh the risks \newline 2: The risks... outweigh the benefits \\
\addlinespace

pph20030 & Q49: Overall, how do you rate the quality of medical care received from your regular doctor in the past 12 months? & 1: Excellent \newline 2: Very good \newline 3: Good \newline 4: Fair \newline 5: Poor \newline 6: Have not seen... \newline 7: Do not have... \\
\addlinespace

pph20031 & Q50: Overall, how satisfied are you with your healthcare coverage? & 1: Very satisfied \newline 2: Moderately satisfied \newline 3: Slightly satisfied \newline 4: Not satisfied \\
\addlinespace

vote2024 & QPID600f: Did you happen to vote in the November 2024 elections for the U.S. President and Congress? & 1: Yes \newline 2: No \\
\addlinespace

candidate2024 & QPID600g: Which candidate did you vote for in the 2024 Presidential election? & 1: Kamala Harris (Democrat) \newline 2: Donald Trump (Republican) \newline 3: Another candidate... \\
\addlinespace

urb\_sub\_rur & Urban/Suburban/Rural & 1: Urban \newline 2: Rural \newline 3: Suburban \\
\addlinespace

ppcm1301 & GOVEMP1: Employer type & 1: Government \newline 2: Private-for-profit company \newline 3: Non-profit organization... \newline 4: Self-employed \newline 5: Working in family business \\
\addlinespace

E140 & E140: Which category best describes your level of employment? & 1: Entry level \newline 2: Experienced (non-manager) \newline 3: Manager/Supervisor... \newline 4: Executive... \\
\addlinespace

MilVet\_1 & DOV: Current Service Member & 1: Current Service Member \\
\addlinespace

MilVet\_2 & DOV: Veteran Service Member & 1: Veteran Service Member \\
\addlinespace

MilVet\_3 & DOV: Non-Military & 1: Non-Military \\
\addlinespace

pppa1640 & Q254: Have you ever been a member of the Reserve or National Guard? & 1: Yes \newline 2: No \\
\addlinespace

pppa1648 & Q26: What is your religion? & 1: Catholic \newline 2: Evangelical/Protestant... \newline 3: Jehovah's Witness \newline 4: LDS (Mormon) \newline 5: Jewish \newline 6: Islam/Muslim \newline 7: Orthodox \newline 8: Hindu \newline 9: Buddhist \newline 10: Unitarian \newline 11: Other Christian \newline 12: Other non-Christian \newline 13: No religion \\
\addlinespace

ppp20197 & QEG22: Are you a citizen of the United States? & 1: Yes \newline 2: No \\
\addlinespace

votereg\_now & Are you currently registered to vote in the U.S.? & 1: Yes, registered \newline 3: No, not registered \newline 4: Not sure \newline 5: No, not eligible \\
\addlinespace

ppfs1482 & Q108: Where do you think your credit score falls & 1: Very poor \newline 2: Poor \newline 3: Fair \newline 4: Good \newline 5: Excellent \newline 6: Don't know \\
\addlinespace

pph10001 & Q1: In general, would you say your health is. . .? & 1: Excellent \newline 2: Very good \newline 3: Good \newline 4: Fair \newline 5: Poor \\
\addlinespace

pph11301 & Q25\_1: Are you a caregiver for one or more children under the age of 18? & 1: Yes \newline 2: No \\
\addlinespace

ppp10035 & Q16: In general, how interested are you in politics and public affairs? & 1: Very interested \newline 2: Somewhat interested \newline 3: Slightly interested \newline 4: Not at all interested \\
\addlinespace

ppcm0160 & Q26: Occupation (detailed) in current or main job & 1: Management \newline 2: Business/Financial... \newline (See full list in data source for codes 3-35) \\
\addlinespace

ppfsasset & Q22: Approx total amount of household savings and investable assets? & 1: Under \$25,000 \newline 2: \$25k - \$49,999 \newline 3: \$50k - \$99,999 \newline 4: \$100k - \$249,999 \newline 5: \$250k - \$499,999 \newline 6: \$500k - \$999,999 \newline 7: \$1M - \$1.9M \newline 8: \$2M or more \newline 9: Not sure \\
\addlinespace

ppc21505 & CU40: How concerned are you about providing personal information over the internet? & 1: Not at all concerned \newline 2: Slightly concerned \newline 3: Somewhat concerned \newline 4: Very concerned \\
\addlinespace

E104\_RET & E104: Do any of the following currently describe you? ... [Retired] & 1: Yes \newline 2: No \\
\addlinespace

E104\_STUD & E104: ... [A student] & 1: Yes \newline 2: No \\
\addlinespace

E104\_STAYHOM & E104: ... [A stay-at-home spouse or partner] & 1: Yes \newline 2: No \\
\addlinespace

E104\_INTERN & E104: ... [Unpaid job/internship/volunteer] & 1: Yes \newline 2: No \\
\addlinespace

E104\_FREELANC & E104: ... [Freelancer/independent contractor] & 1: Yes \newline 2: No \\
\addlinespace

pph10301-7 & Q39/Q40: Alcohol consumption (Beer, Wine, Liquor) & 0: No \newline 1: Yes \\
\addlinespace

pph21901-5 & Q100: Vaccine attitudes (Series) & 1: Do not agree \newline 2: Somewhat agree \newline 3: Agree \newline 4: Strongly agree \\
\addlinespace

ppc21607 & CU44: I use social network sites to communicate with others more than email... & 1: Do not agree \newline 2: Somewhat agree \newline 3: Agree \newline 4: Strongly agree \\
\addlinespace

pph1* & Q19/Q19a: Medical/Mental Health Conditions (ADHD, Anxiety, Depression, etc.) & 0: No \newline 1: Yes (Condition present) \\
\addlinespace

ppm2223* & Q9: News Sources (TV, Paper, Internet, Radio, Social Media) & 0: No \newline 1: Yes \\
\addlinespace

ppp20072 & Q27: How often do you attend religious services? & 1: More than once a week \newline 2: Once a week \newline 3: 1-2 times a month \newline 4: Few times a year \newline 5: Once a year or less \newline 6: Never \\
\addlinespace

ppp22210 & Q34: Household gun ownership & 1: Yes \newline 2: No \\
\addlinespace

ppp22211 & Q36: Personal gun ownership & 1: Yes \newline 2: No \\
\addlinespace

ppm22229 & Q10: How closely do you follow politics...? & 1: Very closely \newline 2: Fairly closely \newline 3: Not very closely \newline 4: Not at all closely \\
\addlinespace

vote2020 & Did you happen to vote in the November 2020 elections...? & 1: Yes \newline 2: No \\
\addlinespace

candidate2020 & Which candidate did you vote for in the 2020 Presidential election? & 1: Joe Biden (Democrat) \newline 2: Donald Trump (Republican) \newline 3: Another candidate \\
\addlinespace

partyid7 & DERIVED: Political party affiliation (7 categories) & 1: Strong Rep \newline 2: Not Strong Rep \newline 3: Leans Rep \newline 4: Undecided/Ind/Other \newline 5: Leans Dem \newline 6: Not Strong Dem \newline 7: Strong Dem \\
\addlinespace

ppp10012 & Q11: In general, do you think of yourself as... & 1: Extr. liberal \newline 2: Liberal \newline 3: Slightly liberal \newline 4: Moderate \newline 5: Slightly conservative \newline 6: Conservative \newline 7: Extr. conservative \\
\addlinespace

PPEDUCAT & Education (4 Categories) & 1: No HS diploma \newline 2: HS grad \newline 3: Some college/Assoc \newline 4: Bachelor's or higher \\
\addlinespace

PPETHM & Race / Ethnicity & 1: White, Non-Hisp \newline 2: Black, Non-Hisp \newline 3: Other, Non-Hisp \newline 4: Hispanic \newline 5: 2+ Races, Non-Hisp \\
\addlinespace

PPGENDER & Gender & 1: Male \newline 2: Female \\
\addlinespace

PPREG4 & Region 4 - Based on State of Residence & 1: Northeast \newline 2: Midwest \newline 3: South \newline 4: West \\
\addlinespace

PPRENT & Ownership Status of Living Quarters & 1: Owned \newline 2: Rented \newline 3: Occupied w/o rent \\
\addlinespace

PPMSACAT & MSA Status & 0: Non-Metro \newline 1: Metro \\
\addlinespace

PPEDUC5 & Education (5 Categories) & 1: No HS \newline 2: HS grad \newline 3: Some college \newline 4: Bachelor's \newline 5: Master's or higher \\
\addlinespace

PPINC7 & Household Income & 1: Less than \$10k \newline 2: \$10k-\$24,999 \newline 3: \$25k-\$49,999 \newline 4: \$50k-\$74,999 \newline 5: \$75k-\$99,999 \newline 6: \$100k-\$149,999 \newline 7: \$150k+ \\
\addlinespace

PPMARIT5 & Marital Status & 1: Married \newline 2: Widowed \newline 3: Divorced \newline 4: Separated \newline 5: Never married \\
\addlinespace

PPEMPLOY & Current Employment Status & 1: Full-time \newline 2: Part-time \newline 3: Not working \\
\addlinespace

PPHOUSE4 & Housing Type & 1: 1-family detached \newline 2: Condo/townhouse \newline 3: 2+ apartments \newline 4: Other \\

\end{longtable}

\section{Unpacking response confidence and diversity}\label{secA3}

To further understand the behavioral differences between demographic persona and SPIRIT persona
conditioning, we conduct a set of diagnostic analyses focusing on response confidence,
answer diversity, and their relationship to inference accuracy.
All analyses in this section are conducted on GPT-5-mini, which represents a stable
and a well-performing model in the main experiments.

\subsection{Response confidence distribution}

\begin{figure}[H]
    \centering
    \includegraphics[width=0.8\linewidth]{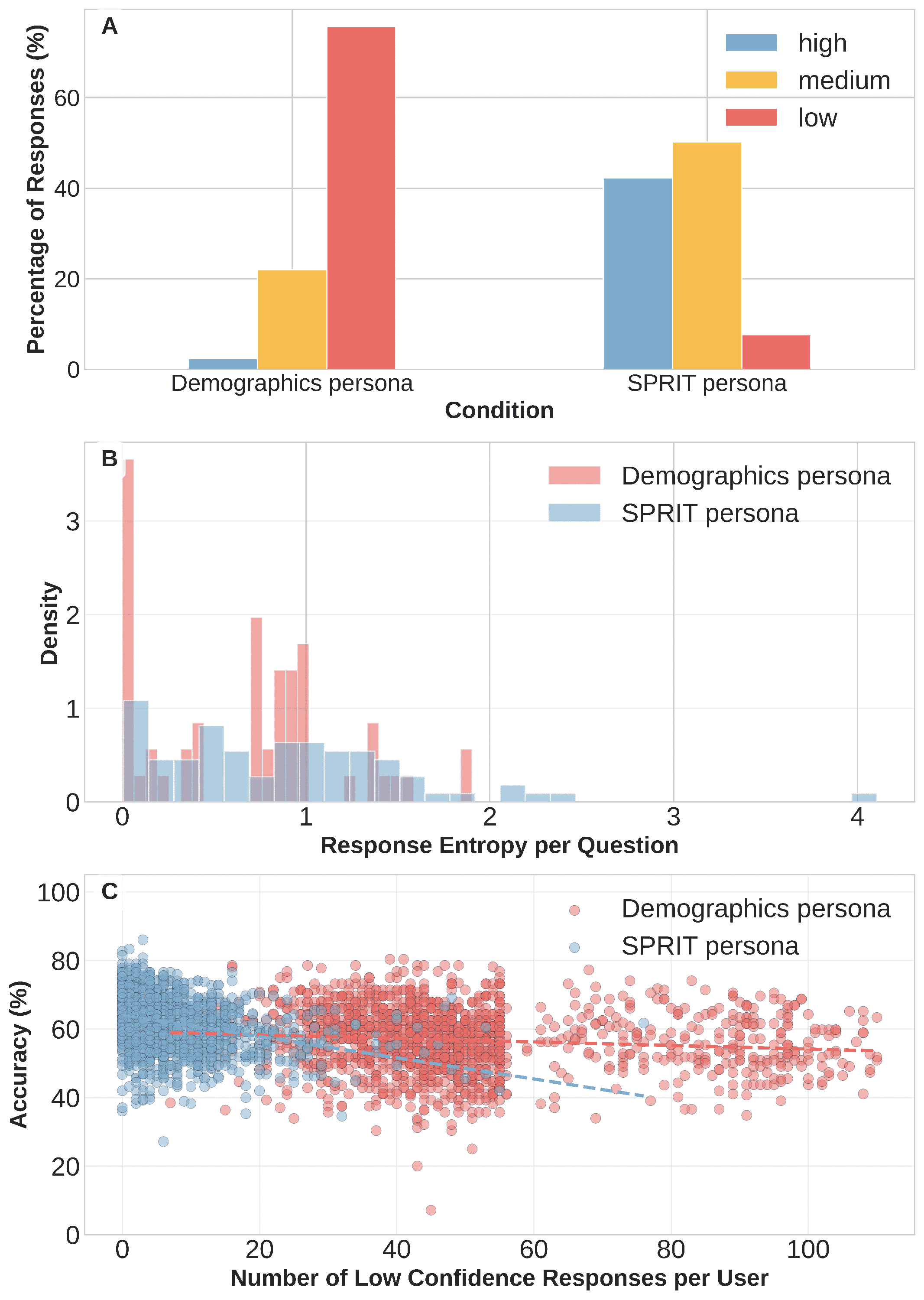}
    \caption{
    Diagnostic analyses of response confidence and diversity for GPT-5-mini.
    \textbf{A}, Distribution of response-level confidence categories under
    demographic persona and SPIRIT persona conditioning.
    \textbf{B}, Distribution of response entropy per question, where lower entropy
    indicates more fixed or biased response tendencies.
    \textbf{C}, Relationship between user-level accuracy and the number of
    low-confidence responses per user, shown separately for each condition with
    linear trend lines.
    }
    \label{fig:appendix-confidence-diversity}
\end{figure}

Figure~\ref{fig:appendix-confidence-diversity}A compares the distribution of response-level
confidence categories across conditions.
Demographic persona conditioning is dominated by low-confidence responses,
whereas SPIRIT persona conditioning produces substantially higher proportions of
medium- and high-confidence responses.

This pattern is expected.
Under demographic persona conditioning, inference relies on sparse, population-level
signals that are often insufficient to support confident predictions at the individual level.
By contrast, SPIRIT persona conditioning incorporates non-demographic attributes inferred
from user-generated text, providing richer contextual grounding and enabling more
confident responses.

\subsection{Response diversity and answer tendency}

We next examine response diversity using entropy computed at the question level
(Figure~\ref{fig:appendix-confidence-diversity}B). To quantify response diversity, we compute Shannon entropy at the question level.
For each question, we collect the inferred categorical responses across users
and estimate the empirical response distribution.
Let $p(v)$ denote the proportion of users assigned response value $v$ for a given question.
Response entropy is defined as
\[
H = - \sum_{v} p(v) \log_2 p(v),
\]
where higher entropy indicates greater variability in inferred responses,
and lower entropy reflects more concentrated or deterministic response patterns.
Entropy is computed only for questions with more than one valid inferred response. Lower entropy indicates more fixed or biased response tendencies,
whereas higher entropy reflects greater variability across users.

Demographic persona conditioning exhibits substantially lower response entropy,
indicating a strong tendency toward fixed or default answers.
In practice, this manifests as systematic preference for negative responses
(e.g., answering ``No'' to behavioral or health-related questions),
which compresses the range of possible inferred values.
In contrast, SPIRIT persona conditioning yields higher entropy distributions,
suggesting that persona-based inference better preserves heterogeneity
across users and questions.

\subsection{Low-confidence responses and user-level accuracy}

Finally, we analyze the relationship between inference accuracy and the number of
low-confidence responses per user (Figure~\ref{fig:appendix-confidence-diversity}C).
Under demographic persona conditioning, users with larger numbers of low-confidence
responses exhibit a clearer degradation in accuracy.
This trend reflects the cumulative effect of uncertain, weakly grounded inferences.

By comparison, SPIRIT persona conditioning substantially reduces the prevalence of
low-confidence responses and weakens the negative association between low confidence
and accuracy.
This suggests that persona-based inference not only improves overall performance
but also stabilizes reasoning at the user level by reducing reliance on uncertain guesses.

These analyses clarify why demographic persona conditioning can appear
competitively accurate on some attributes while still producing degenerate
response patterns.
Demographic persona inference tends to rely on population-level shortcuts,
leading to low-confidence, low-diversity predictions.
SPIRIT persona conditioning mitigates these issues by grounding inference in
non-demographic attributes, resulting in more confident, diverse,
and behaviorally plausible responses.

\section{Subtentive questions}\label{secA4}
\subsection{Epstein files questions}

The following questions (Table~\ref{tab:ext_epstein}) are used to measure public attitudes toward the release of
the Jeffrey Epstein investigation files and potential involvement of public figures.
Question wording closely follows contemporary public opinion surveys conducted
in late 2025 to early 2026.

\begin{table}[h]
\centering
\caption{Epstein files survey questions and response options}
\label{tab:ext_epstein}
\begin{tabular}{p{0.2\linewidth} p{0.6\linewidth}}
\toprule
\textbf{Question ID} & \textbf{Question wording and response options} \\
\midrule
EPSTEIN FILES RELEASE &
Should the U.S. government release all of its files from the investigation of Jeffrey Epstein? \\
& (1) Yes \quad (2) No \quad (98) Not sure \\[0.5em]

TRUMP EPSTEIN INVOLVEMENT &
Do you think that Donald Trump was involved in crimes allegedly committed by Jeffrey Epstein? \\
& (1) Yes \quad (2) No \quad (98) Not sure \\
\bottomrule
\end{tabular}
\end{table}
\subsection{Abortion attitude question}

Attitudes toward abortion are measured using a standard four-category item
commonly employed in U.S. public opinion surveys (Table~\ref{tab:ext_abortion}).

\begin{table}[h]
\centering
\caption{Abortion attitude survey question}
\label{tab:ext_abortion}
\begin{tabular}{p{0.2\linewidth} p{0.6\linewidth}}
\toprule
\textbf{Question ID} & \textbf{Question wording and response options} \\
\midrule
ABRTLGL &
Do you think abortion should be: \\
& (1) Legal in all cases \\
& (2) Legal in most cases \\
& (3) Illegal in most cases \\
& (4) Illegal in all cases \\
\bottomrule
\end{tabular}
\end{table}

\subsection{Immigration policy battery}

Attitudes toward immigration are measured using a policy battery.
For each item, respondents are asked how much they favor or oppose
the proposed policy (Table~\ref{tab:ext_immigration}).

\begin{table}[h]
\centering
\caption{Immigration policy battery items and response categories}
\label{tab:ext_immigration}
\begin{tabular}{p{0.25\linewidth} p{0.55\linewidth}}
\toprule
\textbf{Item ID} & \textbf{Policy description} \\
\midrule
BRDER & Improving security along the country’s borders \\
REF & Admitting more civilian refugees from countries where people are trying to escape violence and war \\
SKILL & Legally admitting more high-skilled immigrants \\
DIV & Legally admitting immigrants from all over the world to ensure the nation’s immigrant population is diverse \\
LABOR & Legally admitting immigrants who can fill labor shortages \\
DEPORT & Enforcing mass deportations of immigrants living in the country illegally \\
STUD & Allowing international students who receive a college degree in the U.S. to legally work and stay in the country \\
MARRY & Allowing undocumented immigrants to legally work and stay in the country if they are married to a U.S. citizen \\
\bottomrule
\end{tabular}
\end{table}

\noindent
Response categories for all immigration items are:
(1) Strongly favor,
(2) Somewhat favor,
(3) Somewhat oppose,
(4) Strongly oppose,
(98) Don't know,
(99) Refused.

\subsection{Venezuela military action questions}

The following questions (Table~\ref{tab:ext_venezuela}) probe public attitudes toward U.S. policy and potential
military action involving Venezuela.
These items are adapted from contemporaneous polling instruments.

\begin{table}[h]
\centering
\caption{Venezuela-related survey questions}
\label{tab:ext_venezuela}
\begin{tabular}{p{0.2\linewidth} p{0.6\linewidth}}
\toprule
\textbf{Question ID} & \textbf{Question wording and response options} \\
\midrule
VZ\_Q12 &
Has the Trump administration clearly explained what the U.S. intends to do regarding Venezuela? \\
& (1) Yes \quad (2) No \quad (3) Not sure \\[0.4em]

VZ\_Q13 &
Does the Trump administration need to explain what the U.S. intends to do regarding Venezuela? \\
& (1) Yes \quad (2) No \quad (3) Not sure \\[0.4em]

VZ\_Q14 &
How much of a threat do you think Venezuela is to the United States? \\
& (1) Major threat \quad (2) Minor threat \quad (3) Not a threat \quad (4) Not sure \\[0.4em]

VZ\_Q15 &
Would you approve or disapprove of potential U.S. military action in Venezuela? \\
& (1) Approve \quad (2) Disapprove \quad (3) Not sure \\[0.4em]

VZ\_Q16 &
Should President Trump need congressional approval before taking military action in Venezuela? \\
& (1) Yes \quad (2) No \quad (3) Not sure \\[0.4em]

VZ\_Q17 &
Do you approve or disapprove of the current military attacks on boats suspected of bringing drugs from Venezuela? \\
& (1) Approve \quad (2) Disapprove \quad (3) Not sure \\[0.4em]

VZ\_Q18 &
Should the administration show evidence that there are drugs on the boats being attacked? \\
& (1) Yes \quad (2) No \quad (3) Not sure \\[0.4em]

VZ\_Q19 &
Do you think U.S. military action in Venezuela would decrease the amount of drugs coming into the U.S.? \\
& (1) Yes \quad (2) No change \quad (3) Increase \quad (4) Not sure \\
\bottomrule
\end{tabular}
\end{table}

\section{Effects of Social Media Trace Quality on Persona-Based Simulation Accuracy}
\label{secA5}

This appendix reports additional analyses examining how the quality and informativeness
of social media traces relate to the accuracy of SPIRIT-based persona simulations,
as well as a full category-level breakdown of performance across all 81 survey questions
using GPT-5-Mini.

\subsection{Trace Quantity and Persona Confidence as Proxies for Information Quality}

\begin{figure}[H]
    \centering
    \includegraphics[width=\linewidth]{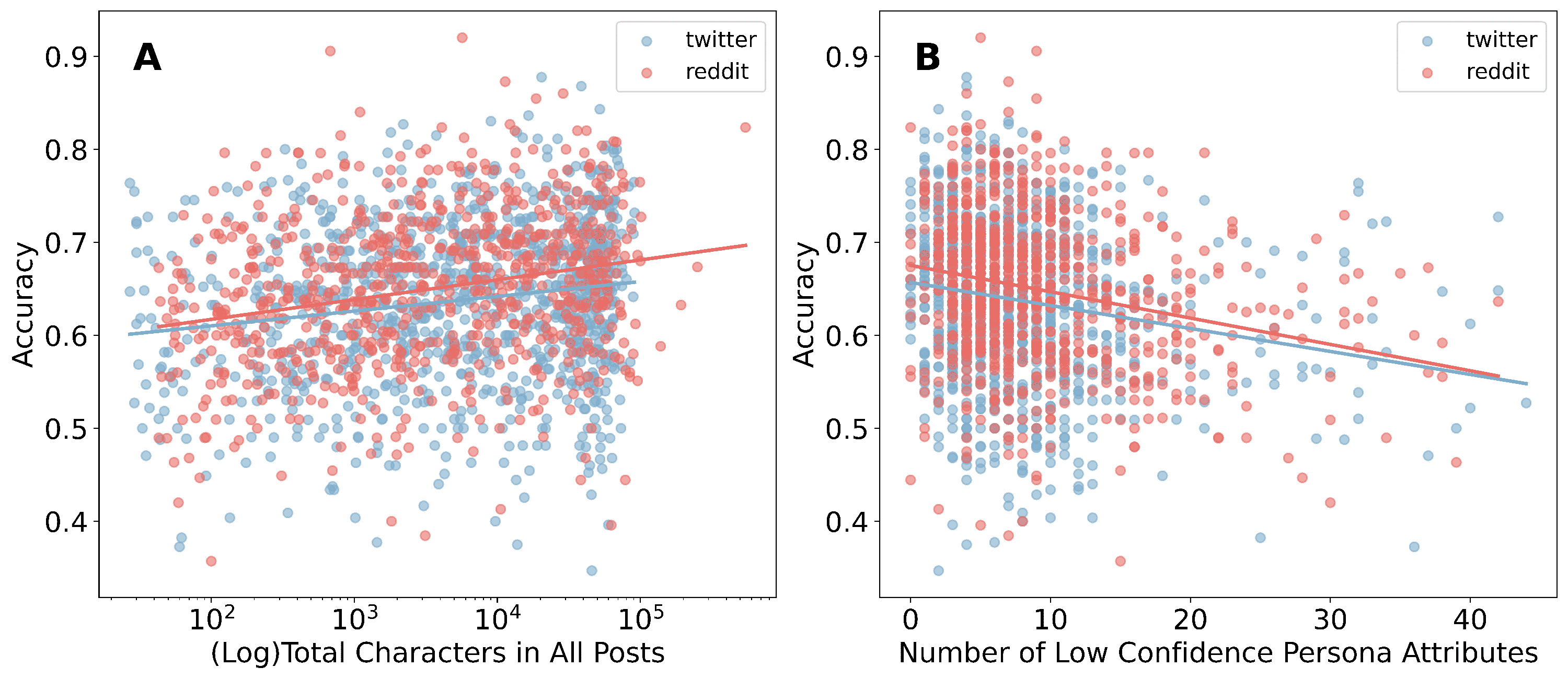}
    \caption{
    \textbf{Relationship between social media trace quality and prediction accuracy.}
    (\textbf{A}) Accuracy as a function of the log-transformed total number of characters
    across all observed posts for each individual.
    (\textbf{B}) Accuracy as a function of the number of low-confidence persona attributes
    inferred by SPIRIT.
    Points represent individual users, colored by platform (Twitter vs.\ Reddit).
    Solid lines indicate linear trends.
    }
    \label{fig:trace_quality}
\end{figure}

Figure~\ref{fig:trace_quality} presents two complementary analyses linking characteristics
of users’ social media traces to downstream prediction accuracy.

Panel~A relates prediction accuracy to the total amount of textual information available
for each individual, measured as the log-transformed total number of characters across all
observed posts. A positive association is observed for both Twitter and Reddit users:
individuals who contribute longer or more extensive social media traces tend to yield more
accurate persona-based predictions. This pattern is consistent with the intuition that
richer behavioral signals allow the persona inference module to estimate latent attributes
more precisely, thereby improving simulation fidelity.

Panel~B examines prediction accuracy as a function of the number of \emph{low-confidence
persona attributes} inferred for a given individual. Each persona dimension produced by
SPIRIT is accompanied by a confidence score; attributes falling below a predefined
threshold are flagged as low-confidence. A clear negative relationship emerges: as the
number of low-confidence attributes increases, simulation accuracy declines. This pattern
holds across platforms and mirrors the result in Panel~A from the opposite direction—when
less information can be reliably inferred from the trace, downstream predictions degrade.

Together, these results show that both trace quantity and persona-level inferential
confidence serve as meaningful proxies for information quality, and that SPIRIT responds
to variation in observational signal in theoretically expected ways.

\subsection{Performance Across Survey Question Categories}
\begin{table}[t]
\centering
\caption{
\textbf{Prediction accuracy by survey question category.}
Reported values are mean exact-match accuracy, standard deviation, number of questions,
total responses, and off-by-one rate for GPT-5-Mini across all 81 survey items.
Categories are ordered from worst to best average accuracy.
}
\label{tab:category_perf}
\begin{tabular}{lccccc}
\hline
\textbf{Category} & \textbf{Avg. Acc.} & \textbf{Std. Dev.} & \textbf{\# Questions} & \textbf{Responses} & \textbf{Off-by-one} \\
\hline
Finances         & 0.245 & 0.430 & 3  & 5{,}185  & 0.407 \\
Technology       & 0.257 & 0.437 & 2  & 3{,}570  & 0.394 \\
Religion         & 0.483 & 0.500 & 2  & 3{,}580  & 0.152 \\
Politics / Media & 0.504 & 0.500 & 11 & 19{,}388 & 0.388 \\
Alcohol          & 0.569 & 0.495 & 8  & 11{,}432 & -- \\
Vaccines         & 0.594 & 0.491 & 7  & 12{,}220 & 0.309 \\
Demographics     & 0.677 & 0.468 & 12 & 21{,}669 & 0.091 \\
Guns             & 0.726 & 0.446 & 2  & 3{,}584  & 0.274 \\
Employment       & 0.738 & 0.440 & 9  & 14{,}752 & 0.129 \\
Health           & 0.764 & 0.425 & 16 & 26{,}610 & 0.341 \\
Voting           & 0.832 & 0.374 & 5  & 8{,}169  & 0.103 \\
Military         & 0.929 & 0.257 & 4  & 6{,}547  & 0.027 \\
\hline
\end{tabular}
\end{table}

To complement the individual-level quality analysis, we report performance disaggregated
by question category across the full set of 81 survey items. Table~\ref{tab:category_perf}
summarizes mean exact-match accuracy, standard deviation, and off-by-one rates for each
category.

Substantial heterogeneity is observed across domains. Categories involving abstract,
private, or infrequently expressed attributes (e.g., finances and technology) exhibit the
lowest accuracy and highest variance. In contrast, domains associated with stable
self-concepts and repeated public expression—such as health, employment, voting, and
military attitudes—show substantially higher accuracy and lower dispersion.

These category-level patterns align with the trace-quality analysis above, reinforcing the
conclusion that persona-based simulation performs best when attitudes are internally
consistent and well-supported by observable discourse.

\subsection{Summary}

Across both individual- and category-level analyses, prediction accuracy scales
systematically with the informativeness of social media traces. Persona confidence scores
provide a useful internal diagnostic of simulation reliability, while category-level
performance reflects theoretically grounded differences in attitude expression and
stability. Together, these results support the use of SPIRIT as a principled framework for
bridging organic digital traces and survey-based measurement.

\subsection{Item-level deviations in the Venezuela questions}
\label{app:venezuela_deviation}

Table~\ref{tab:venezuela_Q14_deviation} and Table~\ref{tab:venezuela_Q19_deviation} compare the weighted distribution of responses from the persona-conditioned virtual panel with contemporaneous CBS poll marginals for two Venezuela-related items. These two items illustrate a recurring pattern: compared to human respondents, LLM-based agents were less likely to select an unqualified, low-engagement option (e.g., \emph{``not a threat''} or \emph{``no change''}) and instead allocated more probability mass to analytically elaborated categories (e.g., \emph{``minor threat''} or \emph{``would increase drugs''}). We interpret this as a \emph{deliberation bias}: when prompted to explain, simulated agents tend to treat questions as requiring causal analysis and internally consistent justification, whereas survey respondents often rely on heuristics or satisficing strategies.

\begin{table}[htbp]
\centering
\caption{Venezuela Q14: perceived threat to the United States. Weighted AI-agent responses versus CBS poll marginals.}
\label{tab:venezuela_Q14_deviation}
\begin{tabular}{lrrr}
\toprule
Response option & Agents (\%) & CBS (\%) & Diff (pp) \\
\midrule
Major threat     & 20.1 & 13  & +7.1 \\
Minor threat     & 79.9 & 48  & +31.9 \\
Not a threat     & 0.0  & 39  & -39.0 \\
\bottomrule
\end{tabular}
\end{table}

For \textbf{Q14} (Table~\ref{tab:venezuela_Q14_deviation}), agents almost never selected \emph{``not a threat''} (0\%), a response that accounts for 39\% in the CBS poll. Instead, agents concentrated on \emph{``minor threat''}. Inspecting the generated rationales suggests that agents tended to (i) acknowledge multiple indirect channels (e.g., regional instability, migration spillovers, transnational crime) while (ii) rejecting the stronger framing of an existential or imminent threat. This combination naturally pushes responses toward \emph{``minor threat''} rather than \emph{``not a threat''}, even when the overall stance is closer to skepticism than alarmism.

\begin{table}[htbp]
\centering
\caption{Venezuela Q19: whether U.S. military action would decrease drugs entering the U.S. Weighted AI-agent responses versus CBS poll marginals.}
\label{tab:venezuela_Q19_deviation}
\begin{tabular}{lrrr}
\toprule
Response option & Agents (\%) & CBS (\%) & Diff (pp) \\
\midrule
Yes, would decrease drugs             & 39.6 & 37 & +2.6 \\
No, would not change amount of drugs  & 11.8 & 56 & -44.2 \\
Would increase drugs                  & 48.6 & 7  & +41.6 \\
\bottomrule
\end{tabular}
\end{table}

For \textbf{Q19} (Table~\ref{tab:venezuela_Q19_deviation}), the binary framing (``decrease'' vs.\ ``not decrease'') yielded a closer match to the poll, but the three-category version shows a substantial redistribution: agents rarely selected \emph{``no change''} and instead shifted heavily toward \emph{``would increase drugs.''} The corresponding rationales frequently invoked systems-style arguments (e.g., displacement/``balloon'' effects, cartel adaptation, governance breakdown, and instability-induced expansion of illicit markets). This reasoning is coherent, but it likely overstates the degree of analytic engagement typical in survey response processes, where respondents may interpret \emph{``no change''} as a satisficing default or an expression of general skepticism about policy efficacy without committing to a backfire mechanism.

\paragraph{Interpretation.}
Taken together, these deviations are consistent with a deliberation bias in LLM-based panels: simulated respondents preferentially construct mechanistic explanations and therefore avoid options that imply categorical dismissal (e.g., \emph{``not a threat''}) or minimal updating (e.g., \emph{``no change''}) (the full reaosning summary can be seen in Appendix~\ref{app:venezuela_reasoning_catalog}). In practice, this implies that matching human marginals may require either (i) explicitly modeling satisficing/low-effort response styles in the Reasoner prompt, or (ii) treating certain response categories as capturing heterogeneous human heuristics that are not well represented by analytic, justification-seeking generation.

\subsection{ Consolidated reasoning patterns by item and response option (Venezuela module)}
\label{app:venezuela_reasoning_catalog}

To aid interpretability, we summarize the \emph{dominant reasoning patterns} produced by the simulated panel for each Venezuela item and each response option. For each question, we group free-text rationales by the chosen answer category and then synthesize the recurring themes into a short paragraph. The goal is not to reproduce verbatim outputs, but to document the qualitative logic that most frequently accompanied each response.

\paragraph{Venezuela Q12: Has the administration clearly explained the reasons for the Venezuela actions?}
\textbf{Yes, has explained clearly.}
Rationales selecting ``Yes'' generally framed the administration’s messaging as clear and direct, emphasizing an unambiguous stance against socialism and a straightforward objective of opposing the Maduro regime. These responses highlighted a consistent rhetorical posture (e.g., firmness, anti-socialist framing, willingness to act) and interpreted any perceived ambiguity as arising from media interpretation rather than from the administration’s communication.

\textbf{No, has not explained clearly.}
Rationales selecting ``No'' converged on the view that the administration relied on broad slogans and posturing without articulating a coherent strategy. Respondents described the approach as vague, reactive, and lacking concrete details about objectives, mechanisms, or end goals. In this framing, the absence of an explicit plan and the perceived volatility of messaging were treated as evidence that the reasons were not clearly explained.

\paragraph{Venezuela Q13: Should the administration explain the reasons to the American people?}
\textbf{Yes, needs to explain.}
Responses selecting ``Yes'' treated potential military involvement as a uniquely consequential decision that requires public justification and democratic accountability. Rationales emphasized constitutional principles, oversight, and the need for clarity about goals, risks, and an exit strategy. Many invoked historical caution regarding past U.S. interventions and rejected vague ideological framing as insufficient for legitimizing force.

\textbf{No, does not need to explain.}
Responses selecting ``No'' primarily appealed to national-security discretion and executive authority. Rationales emphasized that revealing strategy could undermine effectiveness by telegraphing intentions, and portrayed Congress and the media as slow, politicized, or prone to leaks. This cluster valued decisiveness and secrecy over deliberation, framing transparency demands as counterproductive in security contexts.

\paragraph{Venezuela Q14: Is Venezuela a threat to the United States?}
\textbf{Major threat.}
Rationales selecting ``major threat'' framed Venezuela as a proximate national-security risk, frequently citing drug trafficking as the concrete mechanism through which Venezuela could harm U.S. interests. Responses emphasized regional spillovers (migration, criminal networks), geographic proximity, and the possibility of hostile foreign influence. While socialism was often mentioned, it typically served as an explanatory factor for state collapse rather than as the sole basis of threat perception.

\textbf{Minor threat.}
Rationales selecting ``minor threat'' rejected an existential framing while still acknowledging serious problems. Venezuela was described as primarily a humanitarian crisis and a source of indirect regional instability rather than a direct military adversary. Drug trafficking and spillovers were recognized but positioned as contingent and incremental risks. Many explicitly contrasted Venezuela with higher-order geopolitical threats (e.g., major powers), treating the ``major threat'' framing as inflated or politically instrumental.

\textbf{Not a threat.}
In our simulation, this option was rarely selected; consequently, no stable reasoning cluster emerged for ``not a threat'' beyond generic dismissal of relevance to U.S. security.

\paragraph{Venezuela Q15: Do you approve or disapprove of U.S. military action in Venezuela?}
\textbf{Approve.}
Approval rationales were typically conditional and reluctant: respondents expressed general discomfort with war but accepted intervention if it were limited in scope, short in duration, and tied to concrete objectives such as disrupting drug trafficking or removing specific regime actors. ``Quick and decisive'' intervention, rather than prolonged engagement or nation-building, was the dominant constraint.

\textbf{Disapprove.}
Disapproval rationales emphasized the historical track record of U.S. interventions, mission creep, civilian harm, and unintended consequences. Many argued that military force is a poor instrument for complex political and humanitarian crises and preferred non-military alternatives. Even when acknowledging the severity of Venezuela’s situation, respondents viewed intervention as high-risk and low-reward absent a direct, imminent threat.

\paragraph{Venezuela Q16: Should the administration obtain congressional approval before military action?}
\textbf{Yes, needs congressional approval.}
Rationales selecting ``Yes'' strongly emphasized constitutional checks and balances and treated congressional authorization as a basic requirement for initiating war. War-making was framed as qualitatively different from routine executive action and therefore requiring collective deliberation and democratic legitimacy, regardless of partisan preferences.

\textbf{No, does not need congressional approval.}
Rationales selecting ``No'' prioritized speed and executive discretion, portraying Congress as too slow, gridlocked, or prone to politicization for crisis response. This cluster framed unilateral action as necessary for effective leadership, with transparency and deliberation treated as secondary to operational effectiveness in national-security contexts.

\paragraph{Venezuela Q17: Do you approve or disapprove of attacking boats suspected of drug trafficking?}
\textbf{Approve.}
Approval rationales morally foregrounded the harms of drugs (often framed as existential to communities) and treated traffickers as legitimate targets. Many used ends-justify-the-means reasoning, arguing that decisive interdiction deters smuggling and protects U.S. borders, sometimes downplaying uncertainty in identification.

\textbf{Disapprove.}
Disapproval rationales stressed due process, evidentiary standards, and the risk of misidentifying civilians. These responses framed boat attacks as extrajudicial violence with high potential for escalation and norm violation. Suspicion alone was treated as an unacceptably low threshold for lethal force.

\paragraph{Venezuela Q18: Should the administration provide evidence that Venezuela poses a threat before taking action?}
\textbf{Yes, should show evidence.}
This cluster emphasized legitimacy through transparency, arguing that extraordinary force requires demonstrable evidence and public accountability. Concerns about wrongful harm, false positives, and precedent-setting abuses were central. Evidence was treated as a prerequisite to action rather than a post hoc justification.

\textbf{No, does not need to show evidence.}
This cluster justified acting on suspicion by appealing to operational secrecy and trust in military or intelligence expertise. Releasing evidence was portrayed as tactically dangerous (exposing sources and methods) and procedurally impractical in time-sensitive interdiction. The risk of false positives was implicitly accepted as less costly than letting drugs through.

\paragraph{Venezuela Q19: Will military action decrease drugs reaching the United States?}
\textbf{Yes, would decrease drugs.}
Rationales selecting ``Yes'' invoked supply-chain disruption and deterrence: reducing supply increases cost and risk for traffickers and can slow or reduce flow, even if temporarily. These responses treated interdiction as an imperfect but pragmatically beneficial mechanism, emphasizing chokepoints and upstream disruption.

\textbf{No, would not change the amount of drugs.}
Rationales selecting ``no change'' emphasized demand-side constraints and adaptive trafficking networks. This cluster described interdiction as whack-a-mole displacement that shifts routes and methods without meaningfully reducing total flow as long as U.S. demand remains high. Military force was framed as misaligned with the structural drivers of the drug market.

\textbf{Would increase drugs.}
Rationales selecting ``would increase'' extended the adaptive-systems critique into a backfire mechanism. Responses argued that military disruption generates instability, violence, and governance vacuums that strengthen trafficking organizations, diversify routes, and increase incentives through higher prices. The dominant causal chain was: disruption $\rightarrow$ chaos/instability $\rightarrow$ cartel adaptation/expansion $\rightarrow$ equal or higher trafficking volume, often accompanied by historical analogies to past enforcement shocks.

\subsection{Unweighted persona-bank estimates}
\label{app:unweighted}
\begin{figure}[h]
    \centering
    \includegraphics[width=\linewidth]{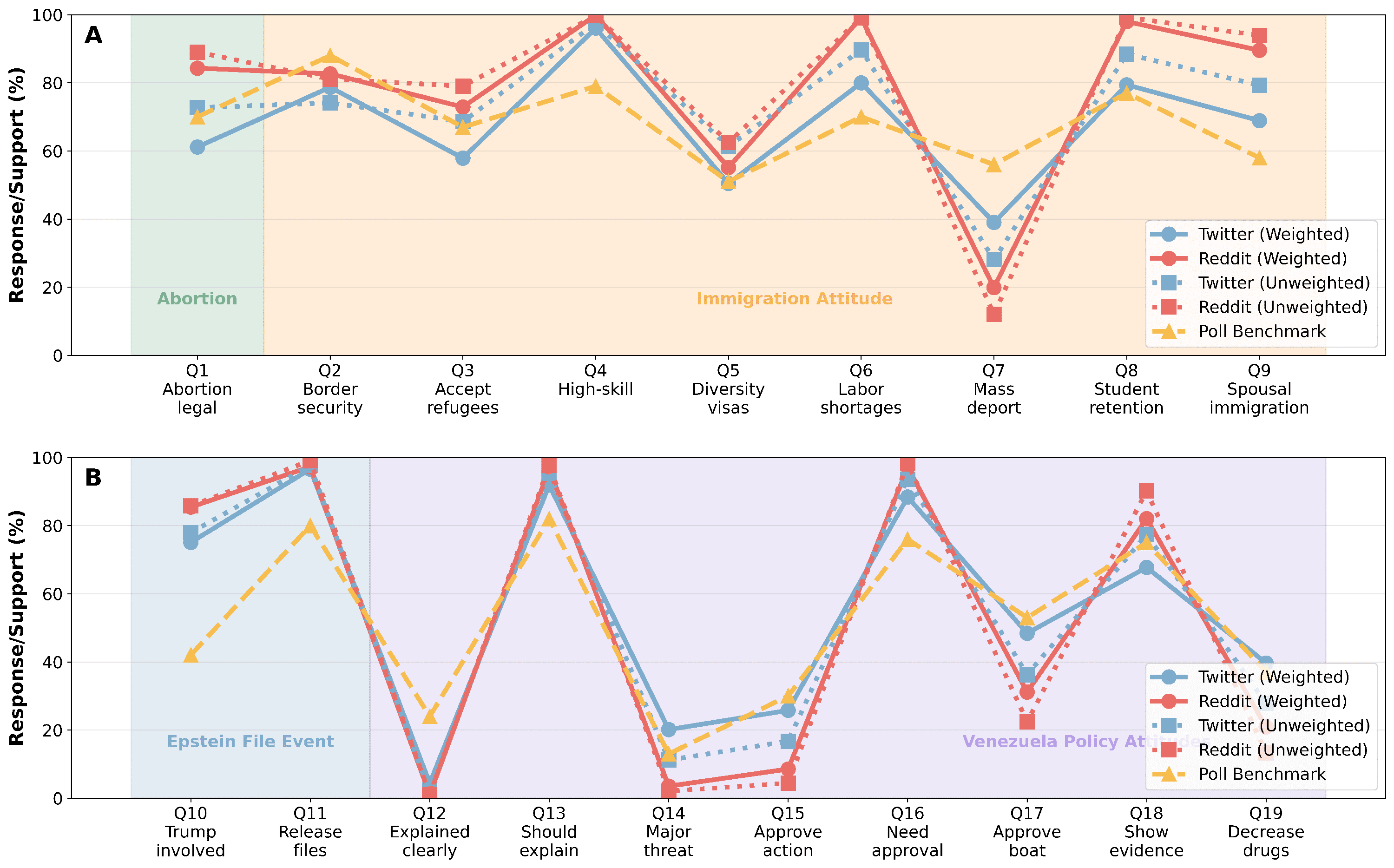}
    \caption{
    Comparison of weighted and unweighted persona-bank responses with polling benchmarks.
    Solid lines denote weighted estimates, while dotted lines denote unweighted aggregates.
    Results are shown for both Twitter- and Reddit-based persona banks across
    long-term attitudinal questions (Panel~A) and event-sensitive questions (Panel~B).
    Unweighted estimates reproduce the same within-cluster directional patterns as weighted estimates,
    but exhibit substantially larger deviations in absolute levels, particularly for Reddit.
    This pattern indicates that weighting primarily improves calibration by
    mitigating selection-induced level bias, while preserving the underlying
    attitudinal structure expressed by the simulated respondents.
    }
    \label{fig:unweighted}
\end{figure}
Figure~\ref{fig:unweighted} reports persona-bank estimates aggregated without weights alongside the weighted results shown in the main text.
This comparison clarifies the role of weighting in the analysis.

Across both panels, unweighted persona-bank responses continue to reproduce
coherent question-to-question structure within each issue-specific cluster.
Items receiving higher support in polling benchmarks are generally ranked higher
by simulated respondents, and lower-support items remain comparatively lower.
This indicates that trend alignment is not an artifact of calibration,
but instead reflects stable attitudinal gradients captured by the inferred personas.

At the same time, unweighted estimates exhibit substantially larger deviations
in absolute levels, particularly for the Reddit-based persona bank.
These deviations are consistent with known compositional imbalances in the underlying
social media samples, including overrepresentation of highly educated and politically
liberal users.
Weighting, therefore, plays a critical role in improving calibration by
reducing selection-induced level bias, rather than altering the qualitative
patterns of opinion expressed by the simulated respondents.

\section{Persona schema}
\label{app:persona_schema}
\subsection{JSON Schema for Painter}
The \textsc{Painter} module is instructed to follow the exact top-level structure below (no additional top-level fields). Field values are constrained to the enumerated options shown.

\begin{small}
\begin{verbatim}
{
  "personality_big5": {
    "openness":       { "approx_level": "...", "confidence": "...", "rationale": "..." },
    "conscientiousness": { "approx_level": "...", "confidence": "...", "rationale": "..." },
    "extraversion":      { "approx_level": "...", "confidence": "...", "rationale": "..." },
    "agreeableness":     { "approx_level": "...", "confidence": "...", "rationale": "..." },
    "neuroticism":       { "approx_level": "...", "confidence": "...", "rationale": "..." }
  },

  "primal_world_beliefs": {
    "good_vs_bad": {
      "value": "leans_good | balanced | leans_bad | unclear",
      "confidence": "high | medium | low",
      "rationale": "..."
    },

    "safe_vs_dangerous": {
      "value": "leans_safe | balanced | leans_dangerous | unclear",
      "confidence": "high | medium | low",
      "rationale": "..."
    },
    "enticing_vs_dull": {
      "value": "leans_enticing | balanced | leans_dull | unclear",
      "confidence": "high | medium | low",
      "rationale": "..."
    },
    "alive_vs_mechanistic": {
      "value": "leans_alive | balanced | leans_mechanistic | unclear",
      "confidence": "high | medium | low",
      "rationale": "..."
    },

    "pleasurable_vs_miserable": {
      "value": "leans_pleasurable | balanced | leans_miserable | unclear",
      "confidence": "high | medium | low",
      "rationale": "..."
    },
    "regenerative_vs_degenerative": {
      "value": "leans_regenerative | balanced | leans_degenerative | unclear",
      "confidence": "high | medium | low",
      "rationale": "..."
    },
    "progressing_vs_declining": {
      "value": "leans_progressing | balanced | leans_declining | unclear",
      "confidence": "high | medium | low",
      "rationale": "..."
    },
    "harmless_vs_threatening": {
      "value": "leans_harmless | balanced | leans_threatening | unclear",
      "confidence": "high | medium | low",
      "rationale": "..."
    },
    "cooperative_vs_competitive": {
      "value": "leans_cooperative | balanced | leans_competitive | unclear",
      "confidence": "high | medium | low",
      "rationale": "..."
    },
    "stable_vs_fragile": {
      "value": "leans_stable | balanced | leans_fragile | unclear",
      "confidence": "high | medium | low",
      "rationale": "..."
    },
    "just_vs_unjust": {
      "value": "leans_just | balanced | leans_unjust | unclear",
      "confidence": "high | medium | low",
      "rationale": "..."
    },

    "interesting_vs_boring": {
      "value": "leans_interesting | balanced | leans_boring | unclear",
      "confidence": "high | medium | low",
      "rationale": "..."
    },
    "beautiful_vs_ugly": {
      "value": "leans_beautiful | balanced | leans_ugly | unclear",
      "confidence": "high | medium | low",
      "rationale": "..."
    },
    "abundant_vs_barren": {
      "value": "leans_abundant | balanced | leans_barren | unclear",
      "confidence": "high | medium | low",
      "rationale": "..."
    },
    "worth_exploring_vs_not_worth_exploring": {
      "value": "leans_worth_exploring | balanced | leans_not_worth_exploring | unclear",
      "confidence": "high | medium | low",
      "rationale": "..."
    },
    "meaningful_vs_meaningless": {
      "value": "leans_meaningful | balanced | leans_meaningless | unclear",
      "confidence": "high | medium | low",
      "rationale": "..."
    },
    "improvable_vs_too_hard_to_improve": {
      "value": "leans_improvable | balanced | leans_too_hard_to_improve | unclear",
      "confidence": "high | medium | low",
      "rationale": "..."
    },
    "funny_vs_not_funny": {
      "value": "leans_funny | balanced | leans_not_funny | unclear",
      "confidence": "high | medium | low",
      "rationale": "..."
    },

    "intentional_vs_unintentional": {
      "value": "leans_intentional | balanced | leans_unintentional | unclear",
      "confidence": "high | medium | low",
      "rationale": "..."
    },
    "needs_me_vs_doesnt_need_me": {
      "value": "leans_needs_me | balanced | leans_doesnt_need_me | unclear",
      "confidence": "high | medium | low",
      "rationale": "..."
    },
    "interactive_vs_indifferent": {
      "value": "leans_interactive | balanced | leans_indifferent | unclear",
      "confidence": "high | medium | low",
      "rationale": "..."
    },

    "interconnected_vs_separable": {
      "value": "leans_interconnected | balanced | leans_separable | unclear",
      "confidence": "high | medium | low",
      "rationale": "..."
    },
    "changing_vs_static": {
      "value": "leans_changing | balanced | leans_static | unclear",
      "confidence": "high | medium | low",
      "rationale": "..."
    },
    "hierarchical_vs_nonhierarchical": {
      "value": "leans_hierarchical | balanced | leans_nonhierarchical | unclear",
      "confidence": "high | medium | low",
      "rationale": "..."
    },
    "understandable_vs_too_hard_to_understand": {
      "value": "leans_understandable | balanced | leans_too_hard_to_understand | unclear",
      "confidence": "high | medium | low",
      "rationale": "..."
    },
    "acceptable_vs_unacceptable": {
      "value": "leans_acceptable | balanced | leans_unacceptable | unclear",
      "confidence": "high | medium | low",
      "rationale": "..."
    }
  },

  "values_and_identities": {
    "salient_identities": [
      {
        "identity": "short label (e.g., parent, gamer, activist, student, etc.)",
        "confidence": "high | medium | low",
        "rationale": "..."
      }
    ],
    "core_values": [
      {
        "value_label": "e.g., equality, order, tradition, autonomy, care, loyalty, etc.",
        "confidence": "high | medium | low",
        "rationale": "..."
      }
    ]
  },

  "life_experiences": {
    "education_and_work": [
      {
        "summary": "...",
        "confidence": "high | medium | low",
        "rationale": "..."
      }
    ],
    "family_and_relationships": [
      {
        "summary": "...",
        "confidence": "high | medium | low",
        "rationale": "..."
      }
    ],
    "turning_points_or_themes": [
      {
        "summary": "Important repeated experiences",
        "confidence": "high | medium | low",
        "rationale": "..."
      }
    ]
  },

  "opinions_and_beliefs": {
    "politics_and_society": [
      {
        "topic": "e.g., elections, immigration, public health, identity politics, etc.",
        "stance_summary": "1–3 sentences summarizing their likely opinion",
        "confidence": "high | medium | low",
        "rationale": "..."
      }
    ],
    "work_and_career": [
      {
        "topic": "work, jobs, academia, gig economy, etc.",
        "stance_summary": "1–3 sentences",
        "confidence": "high | medium | low",
        "rationale": "..."
      }
    ],
    "technology_and_social_media": [
      {
        "topic": "e.g., views on platforms, algorithms, AI, online communities",
        "stance_summary": "1–3 sentences",
        "confidence": "high | medium | low",
        "rationale": "..."
      }
    ],
    "other_recurrent_themes": [
      {
        "topic": "any other recurring domain (e.g., mental health, sports fandom, gaming, religion)",
        "stance_summary": "1–3 sentences",
        "confidence": "high | medium | low",
        "rationale": "..."
      }
    ]
  },

  "interaction_style": {
    "tone": {
      "value": "e.g., sarcastic, earnest, hostile, humorous, supportive, analytical, etc.",
      "confidence": "high | medium | low",
      "rationale": "..."
    },
    "conflict_style": {
      "value": "confrontational | avoidant | accommodating | mixed | unclear",
      "confidence": "high | medium | low",
      "rationale": "..."
    },
    "information_orientation": {
      "value": "news_junkie | casually_informed | low_information | niche_expert | unclear",
      "confidence": "high | medium | low",
      "rationale": "..."
    }
  },

  "meta": {
    "overall_uncertainty_comment": "2–4 sentences",
    "notable_absences": "Brief list or sentence"
  }
}
\end{verbatim}
\end{small}

\subsection{\textsc{Painter} prompt template}
\label{app:painter_prompt}

\noindent\textbf{System prompt.} The following template is used for the \textsc{Painter} module to infer a tentative, probabilistic persona profile from a user’s historical posts.

\begin{small}
\begin{verbatim}
You are an expert computational social scientist trained in survey methodology,
personality psychology, and political behavior.

Your task is to infer a *tentative*, *probabilistic* profile of a single social
media user from their posts.

CRITICAL RULES:
- Use ONLY the textual evidence in the posts (plus ordinary background knowledge
  about language, not about any specific real person).
- Treat all inferences as uncertain hypotheses, not facts.
- When evidence is weak or absent, say "unknown" and explain briefly.
- DO NOT include any demographic fields in your output (age, gender, region, etc.).
  Assume demographics are handled elsewhere in the pipeline.
- Do not attempt to identify or de-anonymize the user, and avoid directly quoting any specific posts; 
  paraphrase them instead.

You must produce TWO parts:

1) A JSON object describing:
   - personality_big5
   - primal_world_beliefs
   - values_and_identities
   - life_experiences
   - opinions_and_beliefs
   - interaction_style
   - meta

2) After the JSON, print a line with exactly:
   ---
   Then output a 2–3 paragraph third-person narrative persona.

JSON SCHEMA (follow this structure exactly; no extra top-level fields):
[see Appendix~\ref{app:persona_schema}]

IMPORTANT OUTPUT FORMAT:
- First, output ONLY the JSON object (no surrounding prose).
- Then a line with exactly three hyphens: ---
- Then the third-person narrative persona (2–3 paragraphs, ~150–300 words).
\end{verbatim}
\end{small}

\subsection{ External survey study prompt templates}
\label{app:external_prompts}

\noindent This appendix documents the prompt templates used in the external survey study. Prompts are shown as \emph{templates}; runtime fields (e.g., \texttt{\{persona\_json\}}, \texttt{\{persona\_narrative\}}, \texttt{\{demographics\}}, \texttt{\{question\_text\}}) are populated programmatically.

\subsubsection{Direct-attitude prompting (no search)}
\label{app:external_prompts_direct}

\begin{small}
\begin{verbatim}
You are participating in a survey. Answer the questions as the person described
below. Use only the information in the persona profile. If the persona provides
no evidence, choose the most plausible option but mark LOW confidence.

PERSONA (JSON):
{persona_json}

PERSONA (Narrative):
{persona_narrative}

DEMOGRAPHICS (provided as context; do not infer new demographics):
{demographics}

INSTRUCTIONS:
- Treat this as a survey response, not a factual exam.
- Give your best answer based on your views/values/habits implied by the persona.
- Do not overthink; avoid long analysis.
- Output a JSON object with one entry per question in the required schema:
  { "value": <int_or_string>, "label": <string>, "confidence": "high|medium|low",
    "reason": <1-3 sentences> }

QUESTION:
{question_text}

RESPONSE OPTIONS:
{options_list}
\end{verbatim}
\end{small}

\subsubsection{Time-sensitive prompting with information acquisition (three-step)}
\label{app:external_prompts_timely}

\paragraph{Step 1: Pre-existing knowledge / prior impressions.}
\begin{small}
\begin{verbatim}
You are participating in a survey about a recent public issue. Answer as the
person described below.

PERSONA (JSON):
{persona_json}

PERSONA (Narrative):
{persona_narrative}

DEMOGRAPHICS (provided as context; do not infer new demographics):
{demographics}

TASK:
Before searching the web, state what you (as this person) already know or
believe about the topic. If you know little or nothing, say so.

OUTPUT (JSON only):
{
  "knowledge_level": "none|minimal|moderate|extensive",
  "what_i_know": "1-4 sentences",
  "where_i_heard_it": "e.g., news, social media, friends, unknown",
  "prior_impression": "1-3 sentences (or 'unknown')"
}

TOPIC:
{topic_name}
\end{verbatim}
\end{small}

\paragraph{Step 2: Persona-conditioned query generation.}
\begin{small}
\begin{verbatim}
Generate 3-5 web search queries that YOU (as this person) would actually type to
learn more about the topic. Queries should reflect your interests, priors, and
trusted sources implied by the persona.

PERSONA (JSON):
{persona_json}

PERSONA (Narrative):
{persona_narrative}

OUTPUT (JSON only):
{ "queries": ["...", "...", "..."] }

TOPIC:
{topic_name}
\end{verbatim}
\end{small}

\paragraph{Step 2b: Summarize retrieved information.}
\begin{small}
\begin{verbatim}
Below are search results (titles/snippets). Summarize what you learned, focusing
on the most relevant points for forming an opinion. Keep it concise.

PERSONA (JSON):
{persona_json}

PERSONA (Narrative):
{persona_narrative}

SEARCH RESULTS:
{search_results}

OUTPUT (JSON only):
{
  "key_points": ["...", "...", "..."],
  "timeframe": "what time period the info seems to cover (if clear)",
  "source_fit": "1-2 sentences on why these sources feel credible or not to you",
  "updated_impression": "1-3 sentences (or 'unchanged')"
}
\end{verbatim}
\end{small}

\paragraph{Step 3: Post-search survey response.}
\begin{small}
\begin{verbatim}
Now answer the survey question as the person described below. Use the persona
AND what you learned from the search summary. This is a survey response, not a
factual exam.

PERSONA (JSON):
{persona_json}

PERSONA (Narrative):
{persona_narrative}

DEMOGRAPHICS:
{demographics}

PRE-KNOWLEDGE (before search):
{preknowledge_json}

SEARCH SUMMARY:
{search_summary_json}

INSTRUCTIONS:
- Choose the option that best matches your view.
- Do not write a long essay; 1-3 sentences of justification is enough.
- Output JSON only in the required schema:
  { "value": <int_or_string>, "label": <string>,
    "confidence": "high|medium|low",
    "reason": <1-3 sentences>,
    "influenced_by_search": true|false }

QUESTION:
{question_text}

RESPONSE OPTIONS:
{options_list}
\end{verbatim}
\end{small}




\end{appendices}


\bibliography{sn-bibliography}

@article{argyle_out_2023,
    title = {Out of {One}, {Many}: {Using} {Language} {Models} to {Simulate} {Human} {Samples}},
    volume = {31},
    issn = {1047-1987, 1476-4989},
    shorttitle = {Out of {One}, {Many}},
    url = {https://www.cambridge.org/core/journals/political-analysis/article/out-of-one-many-using-language-models-to-simulate-human-samples/035D7C8A55B237942FB6DBAD7CAA4E49},
    doi = {10.1017/pan.2023.2},
    language = {en},
    number = {3},
    urldate = {2024-10-20},
    journal = {Political Analysis},
    author = {Argyle, Lisa P. and Busby, Ethan C. and Fulda, Nancy and Gubler, Joshua R. and Rytting, Christopher and Wingate, David},
    month = jul,
    year = {2023},
    pages = {337--351},
}

@misc{horton_large_2023,
    title = {Large {Language} {Models} as {Simulated} {Economic} {Agents}: {What} {Can} {We} {Learn} from {Homo} {Silicus}?},
    shorttitle = {Large {Language} {Models} as {Simulated} {Economic} {Agents}},
    url = {http://arxiv.org/abs/2301.07543},
    doi = {10.48550/arXiv.2301.07543},
    urldate = {2025-12-11},
    publisher = {arXiv},
    author = {Horton, John J.},
    month = jan,
    year = {2023},
}

@misc{park_generative_2024,
    title = {Generative {Agent} {Simulations} of 1,000 {People}},
    url = {http://arxiv.org/abs/2411.10109},
    doi = {10.48550/arXiv.2411.10109},
    urldate = {2025-08-29},
    publisher = {arXiv},
    author = {Park, Joon Sung and Zou, Carolyn Q. and Shaw, Aaron and Hill, Benjamin Mako and Cai, Carrie and Morris, Meredith Ringel and Willer, Robb and Liang, Percy and Bernstein, Michael S.},
    month = nov,
    year = {2024},
}

@misc{yang_oasis_2025,
    title = {{OASIS}: {Open} {Agent} {Social} {Interaction} {Simulations} with {One} {Million} {Agents}},
    shorttitle = {{OASIS}},
    url = {http://arxiv.org/abs/2411.11581},
    doi = {10.48550/arXiv.2411.11581},
    urldate = {2025-11-06},
    publisher = {arXiv},
    author = {Yang, Ziyi and Zhang, Zaibin and Zheng, Zirui and Jiang, Yuxian and Gan, Ziyue and Wang, Zhiyu and Ling, Zijian and Chen, Jinsong and Ma, Martz and Dong, Bowen and Gupta, Prateek and Hu, Shuyue and Yin, Zhenfei and Li, Guohao and Jia, Xu and Wang, Lijun and Ghanem, Bernard and Lu, Huchuan and Lu, Chaochao and Ouyang, Wanli and Qiao, Yu and Torr, Philip and Shao, Jing},
    month = mar,
    year = {2025},
}

@article{wang_large_2025,
    title = {Large language models that replace human participants can harmfully misportray and flatten identity groups},
    volume = {7},
    copyright = {2025 The Author(s), under exclusive licence to Springer Nature Limited},
    issn = {2522-5839},
    url = {https://www.nature.com/articles/s42256-025-00986-z},
    doi = {10.1038/s42256-025-00986-z},
    language = {en},
    number = {3},
    urldate = {2025-12-11},
    journal = {Nature Machine Intelligence},
    author = {Wang, Angelina and Morgenstern, Jamie and Dickerson, John P.},
    month = mar,
    year = {2025},
    pages = {400--411},
}

@misc{ge_scaling_2025,
    title = {Scaling {Synthetic} {Data} {Creation} with 1,000,000,000 {Personas}},
    url = {http://arxiv.org/abs/2406.20094},
    doi = {10.48550/arXiv.2406.20094},
    urldate = {2025-12-11},
    publisher = {arXiv},
    author = {Ge, Tao and Chan, Xin and Wang, Xiaoyang and Yu, Dian and Mi, Haitao and Yu, Dong},
    month = may,
    year = {2025},
}

@inproceedings{murthy_one_2025,
    address = {Albuquerque, New Mexico},
    title = {One fish, two fish, but not the whole sea: {Alignment} reduces language models' conceptual diversity},
    isbn = {979-8-89176-189-6},
    shorttitle = {One fish, two fish, but not the whole sea},
    url = {https://aclanthology.org/2025.naacl-long.561/},
    doi = {10.18653/v1/2025.naacl-long.561},
    urldate = {2025-12-11},
    booktitle = {Proceedings of the 2025 {Conference} of the {Nations} of the {Americas} {Chapter} of the {Association} for {Computational} {Linguistics}: {Human} {Language} {Technologies} ({Volume} 1: {Long} {Papers})},
    publisher = {Association for Computational Linguistics},
    author = {Murthy, Sonia Krishna and Ullman, Tomer and Hu, Jennifer},
    editor = {Chiruzzo, Luis and Ritter, Alan and Wang, Lu},
    month = apr,
    year = {2025},
    pages = {11241--11258},
}

@article{dillion_can_2023,
    title = {Can {AI} language models replace human participants?},
    volume = {27},
    issn = {1364-6613},
    url = {https://www.sciencedirect.com/science/article/pii/S1364661323000980},
    doi = {10.1016/j.tics.2023.04.008},
    number = {7},
    urldate = {2025-12-11},
    journal = {Trends in Cognitive Sciences},
    author = {Dillion, Danica and Tandon, Niket and Gu, Yuling and Gray, Kurt},
    month = jul,
    year = {2023},
    pages = {597--600},
}

@article{kang_deep_2025,
    title = {Deep {Binding} of {Language} {Model} {Virtual} {Personas}: a {Study} on {Approximating} {Political} {Partisan} {Misperceptions}},
    language = {en},
    author = {Kang, Minwoo and Moon, Suhong and Lee, Seung Hyeong and Raj, Ayush and Suh, Joseph and Chan, David M},
    year = {2025},
}

@misc{moon_virtual_2024,
    title = {Virtual {Personas} for {Language} {Models} via an {Anthology} of {Backstories}},
    url = {http://arxiv.org/abs/2407.06576},
    doi = {10.48550/arXiv.2407.06576},
    urldate = {2025-12-11},
    publisher = {arXiv},
    author = {Moon, Suhong and Abdulhai, Marwa and Kang, Minwoo and Suh, Joseph and Soedarmadji, Widyadewi and Behar, Eran Kohen and Chan, David M.},
    month = nov,
    year = {2024},
}

@book{lohr_sampling_2021,
    address = {Boca Raton},
    edition = {3},
    title = {Sampling: {Design} and {Analysis}},
    isbn = {978-0-429-29889-9},
    shorttitle = {Sampling},
    url = {https://www.taylorfrancis.com/books/9780429298899},
    language = {en},
    urldate = {2025-12-11},
    publisher = {Chapman and Hall/CRC},
    author = {Lohr, Sharon L.},
    month = oct,
    year = {2021},
    doi = {10.1201/9780429298899},
}

@misc{zhang_personalizing_2018,
    title = {Personalizing {Dialogue} {Agents}: {I} have a dog, do you have pets too?},
    shorttitle = {Personalizing {Dialogue} {Agents}},
    url = {http://arxiv.org/abs/1801.07243},
    doi = {10.48550/arXiv.1801.07243},
    urldate = {2026-01-26},
    publisher = {arXiv},
    author = {Zhang, Saizheng and Dinan, Emily and Urbanek, Jack and Szlam, Arthur and Kiela, Douwe and Weston, Jason},
    month = sep,
    year = {2018},
}

@article{mairesse_using_2007,
    title = {Using {Linguistic} {Cues} for the {Automatic} {Recognition} of {Personality} in {Conversation} and {Text}},
    volume = {30},
    copyright = {Copyright (c)},
    issn = {1076-9757},
    url = {https://jair.org/index.php/jair/article/view/10520},
    doi = {10.1613/jair.2349},
    language = {en},
    urldate = {2026-01-26},
    journal = {Journal of Artificial Intelligence Research},
    author = {Mairesse, F. and Walker, M. A. and Mehl, M. R. and Moore, R. K.},
    month = nov,
    year = {2007},
    pages = {457--500},
}

@article{clifton_primal_2019,
    address = {US},
    title = {Primal world beliefs},
    volume = {31},
    issn = {1939-134X},
    doi = {10.1037/pas0000639},
    number = {1},
    journal = {Psychological Assessment},
    publisher = {American Psychological Association},
    author = {Clifton, Jeremy D. W. and Baker, Joshua D. and Park, Crystal L. and Yaden, David B. and Clifton, Alicia B. W. and Terni, Paolo and Miller, Jessica L. and Zeng, Guang and Giorgi, Salvatore and Schwartz, H. Andrew and Seligman, Martin E. P.},
    year = {2019},
    pages = {82--99},
}

@article{jost_political_2012,
    address = {Germany},
    title = {Political ideology as motivated social cognition: {Behavioral} and neuroscientific evidence},
    volume = {36},
    issn = {1573-6644},
    shorttitle = {Political ideology as motivated social cognition},
    doi = {10.1007/s11031-011-9260-7},
    number = {1},
    journal = {Motivation and Emotion},
    publisher = {Springer},
    author = {Jost, John T. and Amodio, David M.},
    year = {2012},
    pages = {55--64},
}

@article{schwartz_personality_2013,
    title = {Personality, gender, and age in the language of social media: the open-vocabulary approach},
    volume = {8},
    issn = {1932-6203},
    shorttitle = {Personality, gender, and age in the language of social media},
    url = {https://dx.plos.org/10.1371/journal.pone.0073791},
    doi = {10.1371/journal.pone.0073791},
    language = {en},
    number = {9},
    urldate = {2024-02-15},
    journal = {PLoS One},
    author = {Schwartz, H. Andrew and Eichstaedt, Johannes C. and Kern, Margaret L. and Dziurzynski, Lukasz and Ramones, Stephanie M. and Agrawal, Megha and Shah, Achal and Kosinski, Michal and Stillwell, David and Seligman, Martin E. P. and Ungar, Lyle H.},
    editor = {Preis, Tobias},
    month = sep,
    year = {2013},
    pages = {e73791},
}

@article{ferrara_should_2023,
    title = {Should {ChatGPT} be {Biased}? {Challenges} and {Risks} of {Bias} in {Large} {Language} {Models}},
    issn = {1396-0466},
    shorttitle = {Should {ChatGPT} be {Biased}?},
    url = {http://arxiv.org/abs/2304.03738},
    doi = {10.5210/fm.v28i11.13346},
    urldate = {2026-01-26},
    journal = {First Monday},
    author = {Ferrara, Emilio},
    month = nov,
    year = {2023},
}

@misc{mohsin_fundamental_2025,
    title = {On the {Fundamental} {Limits} of {LLMs} at {Scale}},
    url = {http://arxiv.org/abs/2511.12869},
    doi = {10.48550/arXiv.2511.12869},
    urldate = {2026-01-26},
    publisher = {arXiv},
    author = {Mohsin, Muhammad Ahmed and Umer, Muhammad and Bilal, Ahsan and Memon, Zeeshan and Qadir, Muhammad Ibtsaam and Bhattacharya, Sagnik and Rizwan, Hassan and Gorle, Abhiram R. and Kazmi, Maahe Zehra and Mohsin, Ayesha and Rafique, Muhammad Usman and He, Zihao and Mehta, Pulkit and Jamshed, Muhammad Ali and Cioffi, John M.},
    month = nov,
    year = {2025},
}

@book{kish_survey_1995,
    address = {New York},
    series = {A {Wiley} {Interscience} {Publication}},
    title = {Survey sampling},
    isbn = {978-0-471-10949-5 978-0-471-48900-9},
    language = {eng},
    publisher = {Wiley},
    author = {Kish, Leslie},
    year = {1995},
}

@book{little_statistical_2020,
    address = {Hoboken, NJ},
    edition = {3rd edition},
    series = {Wiley series in probability and statistics},
    title = {Statistical analysis with missing data},
    isbn = {978-0-470-52679-8 978-1-119-48226-0 978-1-118-59601-2 978-1-118-59569-5},
    doi = {10.1002/9781119482260},
    language = {eng},
    publisher = {Wiley},
    author = {Little, Roderick J. A. and Rubin, Donald B.},
    year = {2020},
}

@book{sarndal_model_2003,
    address = {New York Berlin Heidelberg},
    edition = {1. softcover print},
    series = {Springer series in statistics},
    title = {Model assisted survey sampling},
    isbn = {978-0-387-40620-6},
    language = {eng},
    publisher = {Springer},
    author = {Särndal, Carl-Erik and Swensson, Bengt and Wretman, Jan Håkan},
    year = {2003},
}

@article{bailey_authentic_2020,
    title = {Authentic self-expression on social media is associated with greater subjective well-being},
    volume = {11},
    copyright = {2020 The Author(s)},
    issn = {2041-1723},
    url = {https://www.nature.com/articles/s41467-020-18539-w},
    doi = {10.1038/s41467-020-18539-w},
    language = {en},
    number = {1},
    urldate = {2026-01-27},
    journal = {Nature Communications},
    publisher = {Nature Publishing Group},
    author = {Bailey, Erica R. and Matz, Sandra C. and Youyou, Wu and Iyengar, Sheena S.},
    month = oct,
    year = {2020},
    pages = {4889},
}

@misc{park_diminished_2023,
    title = {Diminished {Diversity}-of-{Thought} in a {Standard} {Large} {Language} {Model}},
    url = {http://arxiv.org/abs/2302.07267},
    doi = {10.48550/arXiv.2302.07267},
    urldate = {2025-08-22},
    publisher = {arXiv},
    author = {Park, Peter S. and Schoenegger, Philipp and Zhu, Chongyang},
    month = sep,
    year = {2023},
}

@article{kosinski_private_2013,
    title = {Private traits and attributes are predictable from digital records of human behavior},
    volume = {110},
    issn = {0027-8424, 1091-6490},
    url = {https://pnas.org/doi/full/10.1073/pnas.1218772110},
    doi = {10.1073/pnas.1218772110},
    language = {en},
    number = {15},
    urldate = {2024-02-15},
    journal = {Proceedings of the National Academy of Sciences},
    author = {Kosinski, Michal and Stillwell, David and Graepel, Thore},
    month = apr,
    year = {2013},
    pages = {5802--5805},
}

@misc{taylor_orth_bipartisan_nodate,
    title = {Bipartisan majorities want the government to release its {Jeffrey} {Epstein} records {\textbar} {YouGov}},
    url = {https://today.yougov.com/politics/articles/53427-bipartisan-majorities-want-government-to-release-jeffrey-epstein-records-november-15-17-2025-economist-yougov-poll},
    language = {en-us},
    urldate = {2026-02-03},
    author = {{Taylor Orth}},
    year = {2025}
}

@misc{pew_research_center_public_2025,
    title = {Public {Opinion} on {Abortion}},
    url = {https://www.pewresearch.org/religion/fact-sheet/public-opinion-on-abortion/},
    language = {en-US},
    urldate = {2025-10-16},
    journal = {Pew Research Center},
    author = {{Pew Research Center}},
    month = jun,
    year = {2025},
}

@misc{krogstad_trump_2024,
    title = {Trump and {Harris} {Supporters} {Differ} on {Mass} {Deportations} but {Favor} {Border} {Security}, {High}-{Skilled} {Immigration}},
    url = {https://www.pewresearch.org/race-and-ethnicity/2024/09/27/trump-and-harris-supporters-differ-on-mass-deportations-but-favor-border-security-high-skilled-immigration/},
    language = {en-US},
    urldate = {2026-02-03},
    journal = {Pew Research Center},
    author = {Krogstad, Sahana Mukherjee {and} Jens Manuel},
    month = sep,
    year = {2024},
}

@misc{anthony_salvanto_cbs_2025,
    title = {{CBS} {News} poll finds most would oppose {U}.{S}. military action in {Venezuela}, say {Trump} hasn't explained - {CBS} {News}},
    url = {https://www.cbsnews.com/news/poll-venezuela-u-s-military-action-trump/},
    language = {en-US},
    urldate = {2026-02-03},
    author = {{Anthony Salvanto} and {Jennifer De Pinto}},
    month = nov,
    year = {2025},
}

@article{wright_assessing_2026,
    title = {Assessing personality using zero-shot generative {AI} scoring of brief open-ended text},
    copyright = {2026 The Author(s), under exclusive licence to Springer Nature Limited},
    issn = {2397-3374},
    url = {https://www.nature.com/articles/s41562-025-02389-x},
    doi = {10.1038/s41562-025-02389-x},
    language = {en},
    urldate = {2026-02-03},
    journal = {Nature Human Behaviour},
    publisher = {Nature Publishing Group},
    author = {Wright, Aidan G. C. and Ringwald, Whitney R. and Vize, Colin E. and Eichstaedt, Johannes C. and Angstadt, Mike and Taxali, Aman and Sripada, Chandra},
    month = jan,
    year = {2026},
    pages = {1--15},
}

@article{alwin_reliability_1991,
    title = {The {Reliability} of {Survey} {Attitude} {Measurement}: {The} {Influence} of {Question} and {Respondent} {Attributes}},
    volume = {20},
    issn = {0049-1241},
    shorttitle = {The {Reliability} of {Survey} {Attitude} {Measurement}},
    url = {https://doi.org/10.1177/0049124191020001005},
    doi = {10.1177/0049124191020001005},
    language = {EN},
    number = {1},
    urldate = {2026-02-17},
    journal = {Sociological Methods \& Research},
    publisher = {SAGE Publications Inc},
    author = {ALWIN, DUANE F. and KROSNICK, JON A.},
    month = aug,
    year = {1991},
    pages = {139--181},
}

@misc{toubia_twin-2k-500_2025,
    title = {Twin-{2K}-500: {A} dataset for building digital twins of over 2,000 people based on their answers to over 500 questions},
    shorttitle = {Twin-{2K}-500},
    url = {http://arxiv.org/abs/2505.17479},
    doi = {10.48550/arXiv.2505.17479},
    urldate = {2026-02-21},
    publisher = {arXiv},
    author = {Toubia, Olivier and Gui, George Z. and Peng, Tianyi and Merlau, Daniel J. and Li, Ang and Chen, Haozhe},
    month = may,
    year = {2025},
}

@article{krosnickResponseStrategiesCoping1991,
    title = {Response strategies for coping with the cognitive demands of attitude measures in surveys},
    volume = {5},
    issn = {1099-0720},
    url = {https://onlinelibrary.wiley.com/doi/abs/10.1002/acp.2350050305},
    doi = {10.1002/acp.2350050305},
    language = {en},
    number = {3},
    urldate = {2026-02-23},
    journal = {Applied Cognitive Psychology},
    author = {Krosnick, Jon A.},
    year = {1991},
    pages = {213--236},
}

@article{devilleGeneralizedRakingProcedures1993,
    title = {Generalized raking procedures in survey sampling},
    volume = {88},
    issn = {0162-1459},
    url = {https://www.tandfonline.com/doi/abs/10.1080/01621459.1993.10476369},
    doi = {10.1080/01621459.1993.10476369},
    language = {en},
    number = {423},
    urldate = {2025-09-29},
    journal = {Journal of the American Statistical Association},
    publisher = {ASA Website},
    author = {Deville, Jean-Claude and Särndal, Carl-Erik and Sautory, Olivier},
    month = sep,
    year = {1993},
    pages = {1013--1020},
}

@misc{liSimulatingSocietyRequires2025,
    title = {Simulating {Society} {Requires} {Simulating} {Thought}},
    url = {http://arxiv.org/abs/2506.06958},
    doi = {10.48550/arXiv.2506.06958},
    urldate = {2026-02-23},
    publisher = {arXiv},
    author = {Li, Chance Jiajie and Wu, Jiayi and Mo, Zhenze and Qu, Ao and Tang, Yuhan and Zhao, Kaiya Ivy and Gan, Yulu and Fan, Jie and Yu, Jiangbo and Zhao, Jinhua and Liang, Paul and Alonso, Luis and Larson, Kent},
    month = oct,
    year = {2025},
}

@article{converseNatureBeliefSystems2006,
    title = {The nature of belief systems in mass publics (1964)},
    volume = {18},
    issn = {0891-3811, 1933-8007},
    url = {http://www.tandfonline.com/doi/abs/10.1080/08913810608443650},
    doi = {10.1080/08913810608443650},
    language = {en},
    number = {1-3},
    urldate = {2025-11-11},
    journal = {Critical Review},
    author = {Converse, Philip E.},
    month = jan,
    year = {2006},
    pages = {1--74},
}

@article{zaller_simple_1992,
    title = {A {Simple} {Theory} of the {Survey} {Response}: {Answering} {Questions} versus {Revealing} {Preferences}},
    volume = {36},
    issn = {0092-5853},
    shorttitle = {A {Simple} {Theory} of the {Survey} {Response}},
    url = {https://www.jstor.org/stable/2111583},
    doi = {10.2307/2111583},
    number = {3},
    urldate = {2026-03-03},
    journal = {American Journal of Political Science},
    publisher = {[Midwest Political Science Association, Wiley]},
    author = {Zaller, John and Feldman, Stanley},
    year = {1992},
    pages = {579--616},
}

@misc{yao_react_2023,
    title = {{ReAct}: {Synergizing} {Reasoning} and {Acting} in {Language} {Models}},
    shorttitle = {{ReAct}},
    url = {http://arxiv.org/abs/2210.03629},
    doi = {10.48550/arXiv.2210.03629},
    urldate = {2026-03-13},
    publisher = {arXiv},
    author = {Yao, Shunyu and Zhao, Jeffrey and Yu, Dian and Du, Nan and Shafran, Izhak and Narasimhan, Karthik and Cao, Yuan},
    month = mar,
    year = {2023},
}

@book{mcadamsNarrativeIdentity2011,
    address = {New York, NY, US},
    title = {Narrative identity},
    isbn = {978-1-4419-7987-2 978-1-4419-7988-9},
    doi = {10.1007/978-1-4419-7988-9_5},
    language = {en},
    booktitle = {Handbook of identity theory and research, vols. 1 and 2},
    publisher = {Springer Science + Business Media},
    author = {McAdams, Dan P.},
    year = {2011},
    pages = {99--115},
}

\end{document}